\documentclass[11pt]{article}
\usepackage{jheppub}
\usepackage{amsmath,amssymb,amsfonts,graphicx,slashed,color,amsthm,mathtools,upgreek, enumerate, tensor, subfig}
\usepackage[export]{adjustbox}
\usepackage{arydshln}

\allowdisplaybreaks

\DeclareMathOperator{\tr}{tr}
\newcommand{\beq}{\begin{equation}}
\newcommand{\eeq}{\end{equation}}
\newcommand{\beqn}{\begin{eqnarray}}
\newcommand{\eeqn}{\end{eqnarray}}
\newcommand{\cO}{\mathcal{O}}
\newcommand{\cI}{\mathcal{I}}
\newcommand{\cC}{\mathcal{C}}
\newcommand{\cH}{\mathcal{H}}
\newcommand{\cU}{\mathcal{U}}

\usepackage{hyperref}
\usepackage[utf8]{inputenc}
\usepackage[titletoc]{appendix}
\numberwithin{equation}{section} 

\title{Binding Complexity and Multiparty Entanglement 
}

\author[a, b]{Vijay Balasubramanian}
\author[a]{\!, Matthew DeCross}
\author[a]{\!, Arjun Kar}
\author[a]{\!, Onkar Parrikar}

\affiliation[\,a]{David Rittenhouse Laboratory, University of Pennsylvania,\\
209 S.33rd Street, Philadelphia PA, 19104, U.S.A.}
\affiliation[\,b]{Theoretische Natuurkunde, Vrije Universiteit Brussel (VUB), and \\ International Solvay Institutes, Pleinlaan 2, B-1050 Brussels, Belgium.}

\abstract{
We introduce ``binding complexity", a new notion of circuit complexity which quantifies the difficulty of distributing entanglement among multiple parties, each consisting of many local degrees of freedom.  We define binding complexity of a given state as the minimal number of quantum gates that must act between parties to prepare it.  To illustrate the new notion we compute it in a toy model for a scalar field theory, using certain multiparty entangled states which are analogous to configurations that are known in AdS/CFT to correspond to multiboundary wormholes.  Pursuing this analogy, we show that our states can be prepared by the Euclidean path integral in $(0+1)$-dimensional quantum mechanics on graphs with wormhole-like structure.  We compute the binding complexity of our states by adapting the Euler-Arnold approach to Nielsen's geometrization of gate counting, and find a scaling with entropy that resembles a  result for the interior volume of holographic multiboundary wormholes.  We also compute the binding complexity of general coherent states in perturbation theory, and show that for ``double-trace deformations'' of the Hamiltonian the effects resemble expansion of a wormhole interior in holographic theories.
}

\keywords{}

\begin{document}

\maketitle

\parskip=10pt

\section{Introduction} \label{sec:entanglerobust}

The importance of quantum computational complexity in computer science became apparent after Shor \cite{quant-ph/9508027} proved that the quantum circuit model could solve integer factorization in polynomial time. The typical notion of quantum computational complexity counts the minimal number of simple unitary operations needed to reach some target state from a specific initial state. For instance, one may be interested in how hard it is to prepare the (generically entangled) ground state of a given Hamiltonian starting from an initial state which is factorized across all degrees of freedom. This prompts the related question of whether there is a relation between the strength and structure of the entanglement between degrees of freedom in a quantum state and the complexity of preparing that state. In this work, we answer this question in the affirmative for a type of complexity we call \emph{binding complexity} that counts the number of quantum gates acting on multiple parties simultaneously.

A motivating example that the binding complexity might be connected to the strength of entanglement comes from examination of the two inequivalent classes of multiparty entanglement between three qubits \cite{PhysRevA.62.062314}, the GHZ and W states, and their $n$-party generalizations:
\begin{align}
|\psi_{\text{GHZ}}\rangle &= \frac{1}{\sqrt{2}}\left(|00\ldots 0\rangle + |11\ldots 1\rangle \right) \\
|\psi_{\text{W}}\rangle &= \frac{1}{\sqrt{n}}\left(|00\ldots 01\rangle + |00\ldots 10\rangle + \ldots + |10\ldots 00\rangle \right).
\label{Wstate}
\end{align}
The GHZ states are separable upon tracing out any subset of the parties, whereas the W states are not. In this sense, the W states can be thought of as possessing more robust entanglement. We can understand the structure of these states better by computing the entanglement entropy of one qubit with the rest, as the number of qubits $n$ grows large.  We would normally understand this quantity as a diagnostic of the strength of entanglement between parties.

In more detail, the entanglement entropy corresponding to a partition $(A, \bar{A})$ of degrees of freedom in a quantum state is defined as the von Neumann entropy of the reduced density matrix on $A$: $S_A = -\text{Tr} (\rho_A \ln \rho_A)$. For the GHZ states, we find that entanglement entropy of a single qubit with the rest of the system is 
\begin{align}
S_{1, \text{GHZ}} = \ln 2, \label{eq:ghzent}
\end{align}
which is constant, nonzero, and independent of $n$.  By contrast, for the W states the single qubit entropy is 
\begin{align}
S_{1, \text{W}} = -\left(\frac{n-1}{n} \ln \frac{n-1}{n} + \frac{1}{n} \ln \frac{1}{n}\right) \to 0 \quad \text{as} \quad n \to \infty. \label{eq:went}
\end{align}
It is tempting to conclude from (\ref{eq:ghzent}) and (\ref{eq:went}) that the GHZ states possess ``stronger'' entanglement, at least for large $n$, since there is always maximal entanglement between even a single party and the rest. This seems to be in qualitative tension with our conclusion above that the W states have a more robust pattern of entanglement.

However, one should reinterpret these equations using the principle of monogamy of entanglement \cite{PhysRevA.61.052306}: although $S_{1, \text{GHZ}}$ is constant, tracing out one party removes all of the entanglement as the remaining state is separable. Conversely, $S_{1, \text{W}}$ is small because as the number of parties grows large, tracing out one party only removes a very small amount of entanglement: nearly all of the entanglement remains tied up between the remaining $n-1$ qubits which are still approximately in a W state. We would like to define a quantity that captures this sort of ``robustness" of entanglement: it is distributed between several parties and is difficult to destroy. 

Correspondingly, let us consider quantum circuits preparing the GHZ and W states, and the binding complexities associated to them. To compute complexity we must fix a set of allowed gates that we may use to prepare states. Here, we take the gate set to be the set of all one-qubit or two-qubit unitary operators, although typically we will want further restrictions on which unitaries are allowed.

\begin{figure}[htbp!]
\begin{center}
\includegraphics[width=.5\textwidth]{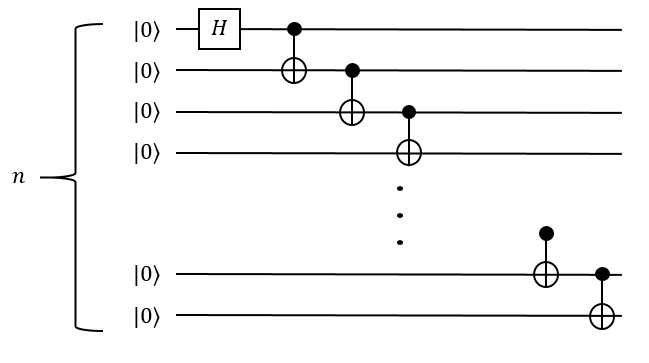}
\caption{Quantum circuit diagram preparing the GHZ state from the factorized state $|0\rangle^{\otimes n}$. The box labeled $H$ indicates the Hadamard operator, a particular unitary one-qubit gate, while the symbol connecting lines refers to the CNOT operator, a unitary two-qubit gate. Here $\mathrm{CNOT} = |0\rangle \langle 0|_A \otimes 1_B +  |1\rangle \langle 1|_A \otimes \sigma^x_B$ and 
$H = |+\rangle \langle 0| + |-\rangle \langle 1|$.
\label{fig:ghzcircuit}}
\end{center}
\end{figure}

In the case of the GHZ states, it is very easy to explicitly write down a circuit that prepares a generalized GHZ state from the factorized state $|0\rangle^{\otimes n}$ (Fig.~\ref{fig:ghzcircuit}). This circuit uses $n$ gates, $n-1$ of which act on multiple parties. The binding complexity, i.e. the number of gates acting on multiple parties at once, is simply $n-1$.  It is easy to understand that one cannot write a more efficient circuit to construct a GHZ state because a minimum of $n-1$ two-party gates are required simply to couple all of the qubits; otherwise, the state will factorize across some partition.

In the case of the W states, it is not simple to write down a circuit, and there is no proof of minimality. However, \cite{1606.09290} gives a deterministic construction of arbitrary W states that requires $\frac{1}{2} n(n+1) - 2 \sim \mathcal{O}(n^2)$ two-qubit gates. To our knowledge no asymptotically more efficient construction has been found. In fact, we should expect that none exists -- intuitively, since the W state is not separable upon tracing out any number of parties, it is as if ${n \choose 2} \sim \mathcal{O}(n^2)$ gates have been used to entangle all pairs of qubits. Consequently, at least in the qubit context, we see that the binding complexity is a natural diagnostic of the robustness of entanglement -- the minimal number of gates required to entangle different parties naturally controls how entangled the parties become in the final state.  Indeed, we will demonstrate bounds relating binding complexity to other measures of robustness such as entanglement negativity, which quantifies non-separability of quantum states.

We will study binding complexity in a toy model of a free scalar field \cite{Jefferson2017}, which reduces to a system of harmonic oscillators.  Binding complexity is defined as the minimum number of gates acting on multiple parties that is needed to prepare the state starting from a specified reference.   In Nielsen's geometric approach \cite{quant-ph/0502070, quant-ph/0603161, quant-ph/0701004} to complexity, one places a Riemannian metric on the space of unitaries, so that complexity is measured by the geodesic distance between the identity and the unitary operator that makes the state of interest.  We choose a metric that is infinitesimal in directions that act only on a single party, so that the geodesic length measures the binding complexity that we want to study.\footnote{The Nielsen approach was previously extended to free fermion fields in \cite{fermion1, fermion2}, coherent states of free scalar fields in \cite{coherent}, states in $\phi^4$ theory in \cite{interacting}, applied to the study of complexity growth following a quench in \cite{quench1, quench2}, and used to study the complexity of Hamiltonians and quantum phase transitions in \cite{hamilcomplex}. An axiomatic study of the Finsler geometry in the Nielsen approach and comparisons to the holographic expectation in thermofield double states and their time evolutions was undertaken in \cite{PhysRevD.97.066004, 1710.00600, 1803.01797, 1809.06678}. Other approaches to field theory definitions of complexity include complexity from the Fubini-Study metric using momentum space \cite{complexityqft2} and complexity from optimization of the Euclidean path integral \cite{bartek, kyoto1, kyoto2, Bhattacharyya2018, 1808.09072, Molina-Vilaplana2018}. Current work has taken first steps towards understanding the Nielsen complexity in CFT and connecting it to the path-integral complexity \cite{Magan2018, 1807.04422}. Most recently, it was argued that the Nielsen complexity is superior to several of the other methods as only the Nielsen complexity displays the correct behavior under certain forward- and backward- time evolutions \cite{complexitytime}.}

A key step in the computation of circuit complexity is the choice of the gate set.  The vacuum wavefunction for a coupled oscillator system is a Gaussian of the schematic form $e^{-\vec{x}^T \Omega \vec{x}}$. In \cite{Jefferson2017}, the gate set acting on such states was chosen to change the components of $\Omega$.  We will divide the oscillators into ``parties'' defined by block structure in $\Omega$. We want to compute the binding complexity of states that are entangled between these parties.  Essentially, this involves only counting the gates from \cite{Jefferson2017} that act across parties -- we will call these the {\it relevant} gates.  To calculate binding complexity we employ the Euler-Arnold approach to simplify the geodesic equation using the Lie algebra of the gate set.\footnote{Nielsen suggested applying the Euler-Arnold equation in \cite{quant-ph/0701004}, which was originally explained in \cite{arnold}. A nice review can be found in \cite{terrytao}.}

It has been suggested that entanglement in quantum field theory can be holographically realized by wormholes between otherwise disconnected regions of spacetime \cite{ERepr, Vijay2014}.  In these contexts complexity in field theory has been conjectured to be dual to the volume or action of an interior region of the wormhole \cite{1406.2678, 1402.5674, 1403.5695, 1408.2823}.  It is also possible in 2+1 dimensions to construct wormholes that connect multiple asymptotic regions \cite{Brill1, Brill2, Brill3, Skenderis2011, Krasnov1, Krasnov2}. Recently these geometries were used to study  multipartite holographic entanglement \cite{Vijay2014,marolf,Fu2018,ross}. Since binding complexity measures the difficulty of entangling the wavefunctions of multiple otherwise disconnected parties, we conjecture that it is related to the interior volume of multiboundary wormholes, i.e., 
$$\text{Binding Complexity = Volume of Wormhole Interior}.$$

We address this conjecture by computing the binding complexity for a natural class of multiparty entangled states in our toy model, and showing that it has a linear dependence on entanglement entropy like the interior volume of the multiboundary wormholes of \cite{Skenderis2011, Krasnov1, Vijay2014}.  The CFT states dual to these wormholes were prepared by the Euclidean path integral on a branched bulk topology \cite{Vijay2014}. Consequently, we consider states in our toy model which are prepared by the Euclidean path integral on certain branched graphs with wormhole-like structure.\footnote{A graphical representation of multiboundary wormholes was similarly put forth in \cite{Skenderis2011}, although their graphs were used purely to represent geometric data regarding how to sew various boundaries together.} We find that such states have binding complexity and entanglement structure that are (a) similar to  properties of the wormhole interior volume, and (b) reminiscent of the bit thread perspective on holographic states.\footnote{Motivation for considering graphs of different topology also comes from the recent work \cite{Fu2018} which examined complexities of formation for wormholes of arbitrary internal topology.}  As a further check on the conjecture we test that adding small double-trace interactions between distinct parties causes binding complexity to increase linearly in the expansion parameter, as expected from the volume increase of holographic wormholes in the Gao-Jafferis-Wall approach \cite{Gao2017}.

\section{Lower bounds}
\label{sec:bounds}

To begin, we will demonstrate some 
elementary lower bounds on binding complexity in terms of other measures of the entanglement structure of a state, such as the entanglement entropy, separability, etc. For simplicity, we will focus on a gate set $\mathcal{G}$ consisting only of one and two-party gates, although our arguments can be generalized to $k$-local gates. 

We begin with the simplest example. Imagine that our Hilbert space can be decomposed into two tensor factors:
\beq
\mathcal{H} = \mathcal{H}_A\otimes \mathcal{H}_{B},
\eeq
where $A$ consists of $N_A$ parties and $B$ consists of the remaining $N_B$ parties. Let $\psi$ be a state in this Hilbert space, and consider a unitary quantum circuit which builds $\psi$ from the reference state $|00\ldots 0\rangle$ 
\beq \label{Circ1}
| \psi \rangle = U_1U_2\cdots U_M |00\ldots 0\rangle,
\eeq
where the $U_i$ are one and two party gates which are allowed within our gate set. Of these gates $U_i$, those which act within $A$ or $B$ do not contribute to the entanglement between $A$ and $B$; only the two-party gates which act across this partition will contribute to the entanglement. Let $n_{AB}$ be the number of such gates which act across the partition. As discussed in the introduction, the binding complexity of the state $\psi$ with respect to the partition $\mathcal{H}_A \otimes \mathcal{H}_B$ is equal to the minimum value $n_{AB}$ in the set $\mathcal{M}_{\psi, \mathcal{G}}$ of all the quantum circuits which construct $\psi$ using the gate set $\mathcal{G}$:
\beq
\mathcal{C}_b (A,B) = \mathrm{min}_{\mathcal{M}_{\psi, \mathcal{G}}}(n_{AB}).
\eeq

\begin{figure}[!h]
\centering
\includegraphics[height=2cm]{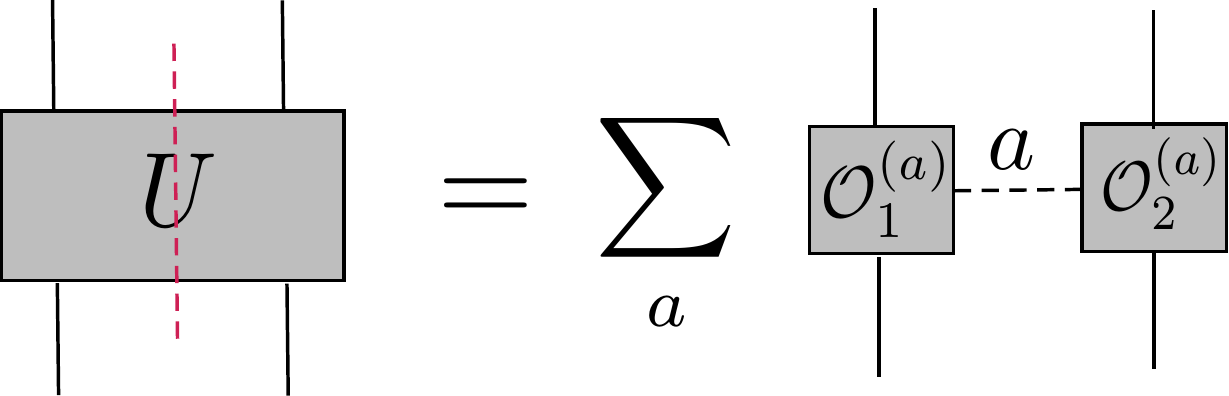}
\caption{\small{We can ``cut'' a two-party gate (denoted by the red dashed line) by using its operator Schmidt decomposition into a sum of products of one-party operators}.\label{fig: cut}}
\end{figure}

In order to study the entanglement structure of $\psi$ given such a quantum circuit in $\mathcal{M}_{\psi, \mathcal{G}}$, we introduce the concept of ``cutting a gate'' (see Fig.~\ref{fig: cut}). Any two-party gate can always be written in the form
\beq
U = \sum_{a=1}^J s_a\cO^{(a)}_1 \otimes \cO^{(a)}_2,
\eeq
where $s_a$ are positive real numbers, the $\{\cO^{(a)}_{1/2}\}$ are a basis of (not necessarily unitary) operators on the first/second party, and $J$ is called the operator Schmidt rank of $U$. This is referred to as operator Schmidt decomposition \cite{2003PhRvA..67e2301N}. Some examples of the operator Schmidt decomposition of two-qubit gates are:
\beq
\mathrm{CNOT} = |0\rangle \langle 0|_A \otimes 1_B +  |1\rangle \langle 1|_A \otimes \sigma^x_B,
\eeq
\beq
\mathrm{SWAP} = \frac{1}{2} \left(1_A\otimes 1_B + \sigma^x_A\otimes \sigma^x_B + \sigma^y_A\otimes \sigma^y_B +\sigma^z_A\otimes \sigma^z_B \right).
\eeq

\begin{figure}[!h]
\centering
\includegraphics[height=3.5cm]{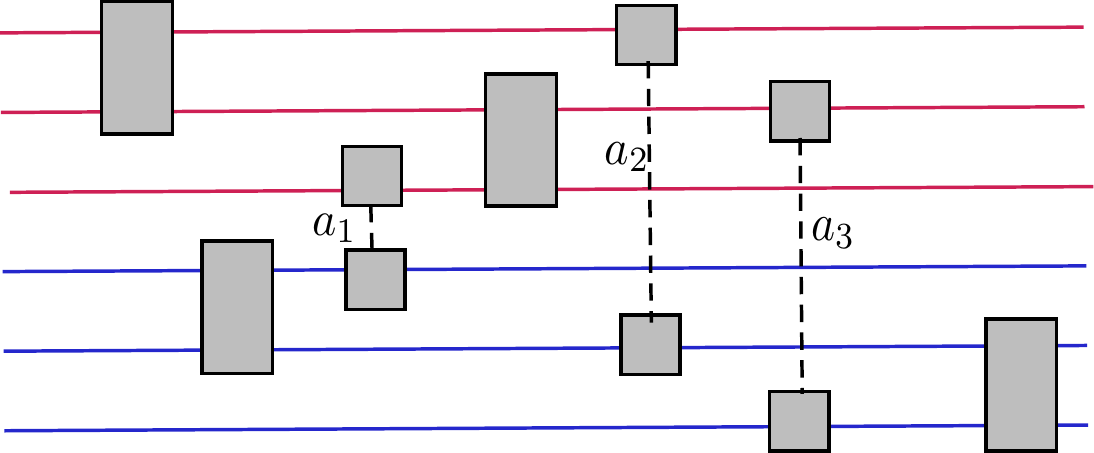}
\caption{\small{A sample piece of a unitary quantum circuit. The red lines denote the subsystem A and the blue lines denote the subsystem B. We have ``cut'' all the two-party gates acting across the bipartition by using their operator Schmidt decomposition into sums of products of one-party operators}.\label{fig: cut2}}
\end{figure}

Turning to our original state $\psi$ in \eqref{Circ1}, we repeatedly employ operator Schmidt decomposition to cut all the two-party gates which act across the partition $\mathcal{H}_A \otimes \mathcal{H}_B$, while leaving all other gates untouched. This allows us to rewrite the state in the form (see Fig.~\ref{fig: cut2})
\beq
|\psi \rangle = \sum_{\vec{a}} p_{\vec{a}} |\psi^A_{\vec{a}}\rangle \otimes |\psi^B_{\vec{a}}\rangle.
\eeq
where $\vec{a} = (a_1,\cdots a_{n_{AB}})$. If we denote by $J_{\mathcal{G}}$ the maximum operator Schmidt rank of any gate in the gate set $\mathcal{G}$, then the above formula shows that the rank of the reduced density matrix on $A$ (or $B$) will be upper bounded by $J_{\mathcal{G}}^{n_{AB}}$. Therefore, the entanglement entropy between $A$ and $B$ satisfies the upper bound\footnote{There is of course the trivial bound on this entropy $S_{A} \leq \ln \mathrm{min}( \mathrm{dim}\,\mathcal{H}_A,\mathrm{dim}\,\mathcal{H}_B)$. However, in general this bound scales with the system size, and will be much weaker than the one in terms of the number of cuts in the quantum circuit.}
\beq
S_{A} \leq \ln(J_{\mathcal{G}})\,n_{AB}.
\eeq
While this upper bound is satisfied by every quantum circuit which constructs $\psi$ from the given gate set $\mathcal{G}$, the bound will be the tightest for the circuit which minimizes $n_{AB}$. Therefore, we conclude that
\beq
S_{A} \leq \ln(J_{\mathcal{G}})\, \mathcal{C}_b(A,B),
\eeq
or equivalently
\beq
\mathcal{C}_b(A,B) \geq \frac{1}{\ln(J_{\mathcal{G}})} S_{A}. 
\eeq
This bound shows that the binding complexity of the state with respect to a bipartition is lower bounded by the entanglement entropy. Intuitively this is clear, because if we are to build a state with a certain amount of entanglement, then we will need sufficiently many gates to achieve this. 

We can easily generalize this bound to multipartite systems. Consider for example a tripartite system $\mathcal{H}_A\otimes \mathcal{H}_{B} \otimes \mathcal{H}_C$ consisting of $N_A$, $N_B$, and $N_C$ qubits respectively. Then by cutting arguments similar to those used above, we obtain
\beq
S_{A} \leq \ln(J_{\mathcal{G}})\,(n_{AB}+ n_{AC}), \;\;S_{B} \leq \ln(J_{\mathcal{G}})\,(n_{BA}+ n_{BC}),\;\;S_{C} \leq \ln(J_{\mathcal{G}})\,(n_{CB}+ n_{CA}), 
\eeq
which gives 
\beq
(S_A + S_B + S_C) \leq 2\ln(J_{\mathcal{G}}) (n_{AB} + n_{BC} + n_{CA}). 
\eeq
For the tripartite system, the binding complexity is defined as the minimum value of $(n_{AB} + n_{BC} + n_{CA})$ across all circuits in $\mathcal{M}_{\psi,\mathcal{G}}$, and therefore we obtain
\beq
\mathcal{C}_b(A,B,C) \geq \frac{1}{\ln(J^2_{\mathcal{G}})} \left(S_A + S_B + S_C \right).
\eeq
Similarly, the $n$-partite generalization of this result is
\beq
\mathcal{C}_b(A_1,\cdots, A_n) \geq \frac{1}{\ln(J^{n-1}_{\mathcal{G}})} \left(S_{A_1} + S_{A_2} +\cdots+ S_{A_n} \right).
\eeq 

So far, we have focused on bounds involving the entanglement entropy. However, as we discussed in the introduction, the entanglement entropy is not always sufficient to probe the fine-grained multiparty entanglement structure of the state. For this purpose, it is useful to consider other information theoretic concepts such as \emph{separability}. Let us consider a tripartite quantum system $\mathcal{H} = \mathcal{H}_A \otimes \mathcal{H}_B \otimes \mathcal{H}_C$. If we trace out $A$, then the reduced density matrix on $BC$ is called \emph{separable} if and only if it can be written in the form
\beq
\rho_{BC} = \sum_i p_i \rho^i_B \otimes \rho^i_C,
\eeq
where $\rho^i_{B/C}$ are density matrices on $B/C$, and $p_i$ are positive real numbers which sum up to 1. In this case, we interpret $\rho_{BC}$ as having no quantum entanglement, i.e., tracing out the subsystem $A$ has destroyed the quantum entanglement between $B$ and $C$. On the other hand, if $\rho_{BC}$ is not separable, the state retains quantum entanglement despite tracing out $A$. A necessary (but not sufficient) criterion for separability is the Peres-Horodecki positivity of partial transpose \cite{1996PhRvL..77.1413P, 1996PhLA..223....1H, 2002PhRvA..65c2314V}. Here, we are instructed to construct the partial transpose $\rho_{BC}^{\Gamma}$ of the density matrix, which is defined as:
\beq
\langle j_B,j_C | \rho_{BC}^{\Gamma}  | \tilde{j}_B, \tilde{j}_C \rangle \equiv \langle \tilde{j}_B,j_C | \rho_{BC}  | j_B, \tilde{j}_C \rangle,
\eeq
where $|j_{B},j_{C}\rangle$ and $|\tilde{j}_{B},\tilde{j}_{C}\rangle$ denote basis vectors for $\mathcal{H}_{B}\otimes \mathcal{H}_C$. If $\rho_{BC}^{\Gamma}$ has any negative eigenvalues, then this necessarily implies that the density matrix $\rho_{BC}$ is not separable. We can therefore quantify the amount of non-separability of $\rho_{BC}$ by the number of negative eigenvalues of the partial transpose $\rho^{\Gamma}_{BC}$. We will denote as $\mathcal{E}_{A|BC}$ the logarithm of one plus the number of negative eigenvalues of $\rho_{BC}^{\Gamma}$. Another measure of the non-separability is the {\it entanglement negativity} $\mathcal{N}_{A|BC}$, which is defined as
\beq
\mathcal{N}_{A|BC}  =\frac{|| \rho^{\Gamma}_{BC} || - 1}{2},
\eeq
where $|| A ||  = \mathrm{Tr}\left(\sqrt{A^{\dagger}A} \right)$ is the trace norm. 

Going back to our state $\psi \in \mathcal{H}_A \otimes \mathcal{H}_B \otimes \mathcal{H}_C$, consider once again some unitary quantum circuit in $\mathcal{M}_{\psi,\mathcal{G}}$ which builds the state from the chosen gate set. By cutting all the two-party gates which act across the tripartition, we can now express the state in the form
\beq
|\psi \rangle = \sum_{\vec{a}, \vec{b}, \vec{c}} p_{\vec{a}, \vec{b}, \vec{c}} |\psi^A_{\vec{a}, \vec{c}}\rangle \otimes |\psi^B_{\vec{a}, \vec{b}}\rangle \otimes |\psi^C_{\vec{b}, \vec{c}}\rangle, 
\eeq
where as before $\vec{a} = (a_1,\cdots a_{n_{AB}})$, $\vec{b} = (b_1,\cdots b_{n_{BC}})$ and $\vec{c} = (c_1,\cdots c_{n_{CA}})$. It is clear from this expression that if we trace out $A$, then the number of negative eigenvalues of $\rho_{BC}^{\Gamma}$ will be upper bounded by the maximum allowed rank of $\rho_{BC}$ minus one (there needs to be at least one positive eigenvalue so the trace can be one), i.e., $(J_{\mathcal{G}}^{n_{AB} + 2n_{BC} + n_{CA}}-1)$. Therefore, 
\beq
\mathcal{E}_{A|BC} \leq \ln (J_{\mathcal{G}}) \left( n_{AB} + 2n_{BC} + n_{CA}\right).
\eeq
Using the same argument by in turn tracing out $B$ and $C$, we obtain
\beq
\left(\mathcal{E}_{A|BC} +\mathcal{E}_{B|CA} + \mathcal{E}_{C|AB}\right)  \leq \ln (J^3_{\mathcal{G}}) \left( n_{AB} + n_{BC} + n_{CA}\right).
\eeq
Once again, the tightest bound is obtained for the circuit which minimizes the right hand side, from which we conclude 
\beq\label{SBound}
\mathcal{C}_b(A,B,C) \geq \frac{1}{\ln (J^3_{\mathcal{G}})} \left(\mathcal{E}_{A|BC} +\mathcal{E}_{B|CA} + \mathcal{E}_{C|AB}\right).
\eeq
At least in the case of qubit systems, the same bound is also true for the (logarithmic) entanglement negativity, i.e., if we replace $\mathcal{E} \to \ln(1+2\mathcal{N})$ in all the terms above. This follows from the fact that the magnitude of all negative eigenvalues is always upper bounded by 1/2 \cite{2013PhRvA..87e4301R}. However, the bound is tighter when stated in terms of $\mathcal{E}$. The bound in equation \eqref{SBound} shows that the binding complexity is a much more fine grained probe of the entanglement structure than the entanglement entropy, and in particular is sensitive to multiparty entanglement measures such as separability.

\section{Computation of the binding complexity}\label{sec:mbycomplex}

To begin, we review the definition of complexity as a geodesic length in the space of unitary operators and explain how the group structure of this space simplifies the geodesic equation. We will keep all sums explicit below as some repeated indices will not be summed. Let us consider a general quantum system with Hilbert space $\cH$. We start by fixing some base state $\psi_0$, such as a completely factorized state. Now consider some other pure state $\psi$ of the entire system, which we wish to study -- for instance, $\psi$ could be the ground state of some interesting Hamiltonian. Let $\cU$ be the space of all unitary maps on $\cH$, and let $\{\cO_I\}$ be a basis for its Lie algebra $\mathfrak{u}$:
\beq\label{StructureConstants}
\left[\cO_I, \cO_J\right] = i\sum_K {f_{IJ}}^K\cO_K.
\eeq
We may think of $\cO_I$ as generators of the elementary unitary gates at our disposal (thus $e^{i\cO_I}$ are the elementary gates). Let $U\in \cU$ be an operator such that
\beq
|\psi\rangle  = U |\psi_0\rangle.
\eeq
In order to define the complexity of $U$, we need a notion of distance on the group manifold $\cU$. One possibility is the standard bi-invariant metric which is obtained from the Cartan-Killing form $K_{IJ}$ on the Lie algebra $\mathfrak{u}$, defined in terms of the structure constants as\footnote{The Cartan-Killing form satisfies $\sum_N {f_{IJ}}^NK_{NM} = -\sum_N {f_{IM}}^NK_{NJ}$, which is simply the statement that it is invariant under adjoint action of the group, i.e., $\mathrm{Tr}_{\text{ad}}([Z,X]Y) + \mathrm{Tr}_{\text{ad}}(X[Z,Y]) = 0$ for any three elements $X,Y,Z$ of the Lie-algebra.\label{CKF}}
\footnote{We have defined the Cartan-Killing form up to overall sign and normalization here, since our main goal is using it to construct a right-invariant Riemannian metric whose normalization is fixed by ``cost factors'' (see \eqref{cg}).}
\beq\label{CKFdef}
K_{IJ} =  \mathrm{Tr}_{\text{ad}}(\cO_I \cO_J) = \sum_{M,N} {f_{IM}}^N {f_{JN}}^M.
\eeq
If we allow gates that can act on any number of qubits at the same time, arbitrarily complex operations could be done in a single step.  Thus, it is necessary to restrict the gate set to be ``local'' in some sense. We will require gates to be ``bilocal'', acting on no more than two qubits at the same time.

So far our discussion has been general, but now we wish to specialize the notion of complexity to study multiparty entanglement. To this end, let us consider a system which has a natural tensor factorization of the form
\beq
\cH = \cH_1 \otimes \cH_2 \otimes \cdots \cH_{N}.
\eeq
In order to study the multiparty entanglement structure (with respect to the above partition) of a state in this Hilbert space, we define the \emph{binding complexity} as the minimal number of gates, required in a quantum circuit construction of $U$, which act on more than one factor at a time, i.e., they act across the chosen partition.  Gates which act within a tensor factor do not add to entanglement, and as such are treated as irrelevant. However, gates which act on two or more factors do contribute to the entanglement between various parties, and as such will be regarded as relevant. We wish to optimize over the number of relevant gates in building the unitary $U$. 

To accomplish this we can define a different inner product $G_{IJ}$ on the Lie algebra, which assigns a different ``cost" for gates acting on one vs. multiple parties. We define the inner product by the metric
\beq\label{cg}
G_{IJ} = \frac{c_I+c_J}{2} K_{IJ}, 
\eeq
where the $c_I$ are the cost factors for the operators $\cO_I$.  We can then construct a  \emph{right-invariant metric} $g$ on $\cU$ as follows: if $X=\frac{dU}{dt}$ is a tangent vector to $\cU$ at some point $U$, then we can define a corresponding Lie algebra element $X U^{-1}$. Then the metric is defined by
\beq\label{RImetric}
g_U (X,Y) = G(X U^{-1}, YU^{-1}).
\eeq

To define the cost factors, let us split our generators $\cO_I$ into $\cO_{\alpha} \in R$ and $\cO_{\bar\alpha}\in \bar{R}$, where $\cO_\alpha$ are the \emph{relevant} generators which simultaneously act on multiple factors, while $\cO_{\bar\alpha}$ are irrelevant and act within individual factors. Then we can simply take the metric $G_{IJ}$ to be of the form in (\ref{cg}) with the cost factors given by
\beq\label{CostFactors}
c_I = \begin{cases} \epsilon^2, &  \;\;\cO_I \in \bar{R}, \\ 1, & \;\; \cO_I \in R, \end{cases} 
\eeq 
where $\epsilon$ is a small parameter that will be taken to zero at the end. We now define the \emph{Nielsen binding complexity} (or simply binding complexity, for brevity) $\cC_b$ of a unitary $U$ as the minimal distance between $U$ and the identity with respect to the above metric, in the limit $\epsilon \to 0$.  Many different unitary operators may prepare the same state -- e.g., we can always multiply one such unitary by others that rotate the part of the Hilbert space that is orthogonal to the reference state. Consequently, the complexity (binding or otherwise) of a \emph{state} as opposed to an operator is defined as the complexity of the simplest unitary operator preparing that state.  In the examples we study it will turn out that there is a unique operator preparing each state, so we can avoid this subtlety.

From this perspective the binding complexity $\cC_b$ of a unitary operator is its minimal geodesic distance from the identity in the metric (\ref{RImetric}) \cite{quant-ph/0502070, quant-ph/0603161, quant-ph/0701004}.\footnote{It was shown in \cite{quant-ph/0502070} that Nielsen's geodesic approach provides a lower bound on gate complexity for an appropriate choice of the inner product $G_{IJ}$. Here we are adopting this approach to compute binding complexity.}  For group manifolds with right-invariant metrics of the form discussed above, the geodesic equation takes a simple form, often referred to as the \emph{Euler-Arnold equation} (perhaps familiar from rigid-body dynamics). Let $U(s)$ be a geodesic on $\cU$, and let $v(s) = \frac{dU}{ds} U^{-1} \in \mathfrak{u} $ be the velocity vector pulled back to the identity. Then the Euler-Arnold equation is
\beq\label{EAeq}
\sum_J G_{IJ} \frac{dv^J}{ds} =   \sum_{L,M,N} {f_{IM}}^LG_{LN} v^Mv^N.
\eeq
There is a slightly different way to express this equation, which will be convenient at times. Let us define a matrix $\cI^I_J$ such that $G_{IJ} = \sum_M K_{IM}\cI^M_J$. If we assume that the Cartan-Killing form is invertible, then we get $\cI^I_J =  \sum_M K^{IM}G_{MJ}$. In terms of $\cI$, the Euler-Arnold equation reads 
\beq
\sum_J {\cI^I}_{J}\frac{dv^J}{ds} = \sum_{L,M,N} {f_{MN}}^I v^M \left({\cI^N}_{L} v^L\right) . \label{c2}
\eeq
where we have used the invariance property of the Cartan-Killing form, explained in footnote \ref{CKF}. Alternatively, if we define $\mathbf{L} =  \sum_{I,J} \cI^I_Jv^J\cO_I$, and $\mathbf{v} = \sum_I v^I\cO_I$, then we obtain
\beq
i \frac{d\mathbf{L}}{ds} = \left[ \mathbf{v}, \mathbf{L}\right].
\eeq
Note that it is crucial that the structure constants mix generators with different cost factors for the term on the right to survive. In order to obtain the geodesics, we must solve equation \eqref{c2} for the velocity $v^I$. We then use this to obtain the geodesic, which satisfies
\beq
\frac{dU}{ds}(s) = i\sum_I v^I(s)\cO_I U(s),
\eeq
and implement the boundary conditions $U(0) = 1$ and $U(1) = U$, where $U$ is the unitary whose complexity we wish to study.

\subsection{Complexity of Gaussian states}\label{sec:HO}

Our starting point is the toy model of \cite{Jefferson2017}, which takes a system of harmonic oscillators as an approximation to a free scalar field theory on an $n$-point lattice. Since we are interested in using this as a setting for the study of multiparty entanglement, we partition the oscillators into $m$ groups of $N$ oscillators each, so that $n = Nm$. We will refer to each group of oscillators as a ``party". The operator content of the theory are the ``position" and ``momentum" operators $\hat{\phi}^i$ and $\hat{\pi}^i$ at each site, with $i = 1, 2, \ldots , n$ and canonical commutation relation $[\hat{\phi}^i, \hat{\pi}^j ] = i\delta^{ij}$. We consider Gaussian states of the form:
\begin{align}
|\Psi \rangle = \left(\frac{\det \Omega}{\pi^n}\right)^{1/4} \int d\vec{\varphi}\: e^{-\frac12 \vec{\varphi}^T \Omega  \vec{\varphi}} |\vec{\varphi} \rangle, \label{eq:gaussian}
\end{align}
where $d\vec{\varphi} = d\varphi^1 \ldots d\varphi^n$, $|\vec{\varphi}\rangle = |\varphi^1\rangle \otimes \ldots \otimes |\varphi^n\rangle$ with $|\varphi^i \rangle$ an eigenstate of $\hat{\phi}^i$, $\Omega$ a symmetric matrix with positive-definite eigenvalues, and the coefficient out front is for normalization.  The vacuum state has a specific $\Omega$.

We will determine the binding complexity of such states with respect to a reference in which $\Omega$ is diagonal.  The gate set for measuring complexity will consist of the Hermitian operators:\footnote{This gate set is universal, i.e. can prepare any state, when we restrict ourselves to the subspace of Gaussian states \eqref{eq:gaussian}. However, it is not sufficient to prepare arbitrary states, for which we would need to supply additional gates.}
\begin{align}
\hat{\cO} (A) = \frac12  \sum_{i,j} A_{ij} (\hat{\phi}^i \hat{\pi}^j + \hat{\pi}^j \hat{\phi}^i).
\end{align}
$A$ is an arbitrary $n \times n$ matrix, so $A \in \mathfrak{gl}(n, \mathbb{R})$. It is straightforward to check that $\bigl[\hat{\cO} (A), \hat{\cO} (B) \bigr] = -i\hat{\cO} ([A,B])$, so the $\hat{\cO} (A)$ operators form a representation of $\mathfrak{gl}(n, \mathbb{R})$. Choosing as generators of $\mathfrak{gl}(n, \mathbb{R})$ the elementary matrices $(M_{ij})_{k\ell} = \delta_{ik} \delta_{j\ell}$, we correspondingly define the generators of the gate set:
\begin{align}
\hat{\cO}_{ij} = \hat{\cO} (M_{ij}) = \frac12 (\hat{\phi}_i \hat{\pi}_j + \hat{\pi}_j \hat{\phi}_i). \label{eq:kops}
\end{align}
(That is, $e^{i\sum_{i,j}\epsilon^{ij}\hat{\cO}_{ij}}$ are the gates we use.) 
A short computation gives the structure constants
\begin{align} \label{SC}
{f_{ij, k\ell}}^{mn} = \delta_{i\ell} \delta_k^m \delta_j^n - \delta_{kj} \delta_i^m \delta_{\ell}^n.
\end{align}
To be clear, the  $\hat{\cO}_{ij}$ are the $\cO_I$ in the discussion above  \eqref{StructureConstants}, where now $I = ij$ is a double-index since we are working with a matrix Lie group.  
A unitary operator that prepares the general Gaussian state (\ref{eq:gaussian}) from the reference state can then be reached from the identity by a continuous sequence of unitary operators, described by the path-ordered exponential
\begin{align}
\hat{U} (s) = \mathcal{P} \exp \left(i \int_0^s ds' \: \sum_{i,j} V^{ij} (s') \hat{\cO}_{ij} \right), \label{eq:unitary}
\end{align}
where $s$ parameterizes a trajectory in the space of unitary operators and the $V^{ij} (s)$ describe the instantaneous direction in the tangent space $\mathfrak{gl}(n, \mathbb{R})$, i.e., ``velocity" in the space of unitary operators.  We pick the boundary condition so that $\hat{U} (1) $ is the unitary operator that prepares the desired state.

To define binding complexity we follow the geodesic formalism described above. In terms of the non-degenerate metric on the space of generators \eqref{cg}, $G_{IJ} \equiv G_{ij,k\ell}$,  operator complexity is defined as the length of the \emph{geodesic} trajectory connecting $\hat{U} (s=1)$ to the identity,
\begin{align}
\mathcal{C} = \int_0^1 ds\: \sqrt{\sum_{i,j,k,\ell} G_{ij, k\ell} V^{ij}(s)  V^{k\ell}(s)}. 
\end{align}
If there are multiple such geodesics,  complexity is defined as the minimum of their lengths.   The relevant and irrelevant operator directions are defined by the ``costs'' in the metric \eqref{CostFactors}, so that 
 $G_{ij,k\ell} = (c_{ij} + c_{k\ell}) K_{ij, k\ell} / 2$.
 Here $c_{ij} = 1$ if $\hat{\mathcal{O}}_{ij} \in R$ and $c_{ij} = \epsilon^2$ if  $\hat{\mathcal{O}}_{ij} \in \bar{R}$
where $R$ is the set of operators $\hat{\mathcal{O}}_{ij}$ such that oscillators $i$ and $j$ are located in different parties. We take $K_{ij,k\ell}$ to be the Cartan-Killing form for $\mathfrak{gl}(n,\mathbb{R})$,
\beq
K_{ij,k\ell} = \left( \delta_{i\ell}\delta_{jk} -\frac{1}{n}\delta_{ij}\delta_{k\ell}\right), \label{eq:ckform}
\eeq
where we have included an additional normalization factor of $\frac{1}{2n}$ for convenience as compared to  \eqref{CKFdef}. In the end, $\epsilon$ will be taken to zero  and is included to make sure that $G$ is non-degenerate. 

A subtlety here is that the Cartan-Killing form for $\mathfrak{gl}(n,\mathbb{R})$ has a degenerate direction, which in our notation reads\footnote{This can equivalently be stated as $\sum_{k,\ell} K_{ij, k\ell} \delta^{k\ell} = 0$, since $\sum_i\hat{\cO_{ii}} = \sum_{ij} \delta^{ij}\hat{\cO}_{ij}$. } $\sum_i\hat{\cO}_{ii}$.  This leads to a degeneracy in the metric, which we had wanted to avoid.  Fortunately, the direction with a vanishing line element is irrelevant (i.e. it represents a gate acting within parties, as opposed to between them).  So the degeneracy does not affect the binding complexity. However it can potentially lead to an ambiguity in the geodesic equation \eqref{EAeq}, because in the degenerate directions the equation becomes $0 = 0$.  Fortunately, in the rigid-body form \eqref{c2}, degeneracies arising from the Cartan-Killing form drop out allowing us to avoid this subtlety.  Other than that, our metric is block diagonal (i.e., does not mix relevant and irrelevant directions), permutation-symmetric between parties, and only the relevant operators creating entanglement between parties contribute to binding complexity.

In the $\epsilon \to 0$ limit, the binding complexity is then 
\begin{align}
\mathcal{C}_b = \int_0^1 ds\: \sqrt{\sum_{\hat{\mathcal{O}}_{ij} \in R} |V^{ij} (s)|^2}. \label{eq:complexity}
\end{align}
To compute the velocities $V^{ij}(s)$ on geodesics we use the Euler-Arnold equation (\ref{c2}).  In the present case this equation takes the form
\begin{align}
\sum_{k,\ell} \cI^{ij}_{k\ell} \frac{dV^{k\ell}}{ds} -  \sum_{k,\ell,p,q,m,n} \cI^{k\ell}_{pq} {f_{mn, k\ell}}^{ij} V^{mn} V^{pq} = 0 ,\label{eq:ea}
\end{align}
where the structure constants are given in \eqref{SC}, and the matrix $\cI^{ij}_{k\ell} = c_{ij} \delta^{i}_k\delta^j_\ell$.\footnote{Even though the Cartan-Killing form is not invertible, it can be checked that solving this equation is equivalent to solving the Euler-Arnold equation \eqref{EAeq}.} To solve (\ref{eq:ea}), we must consider two cases: either $i,j$ are in the same party, or they are in different parties. The resulting equations are
\begin{align}
\epsilon^2 \frac{dV^{ij}}{ds} &= 0,  \qquad\quad  \hat{\mathcal{O}}_{ij} \in \bar{R},\\
\frac{dV^{ij}}{ds} - (1-\epsilon^2) (V^{jj} - V^{ii}) V^{ij} &= 0, \qquad \quad  \hat{\mathcal{O}}_{ij} \in R.
\end{align}
These are in general solved by:
\begin{align}
V^{ij}(s) &= v^{ij},\qquad\qquad\qquad \quad\:\:\: \hat{\mathcal{O}}_{ij} \in \bar{R}, \\
V^{ij} (s) &= v^{ij} e^{(1-\epsilon^2) (v^{jj} - v^{ii})s}, \,\quad  \hat{\mathcal{O}}_{ij} \in R,
\end{align}
where the $v^{ij}$ are integration constants.  We are going to choose final states that are symmetric between the parties just like the initial states.  Thus we expect to find a geodesic that is permutation-symmetric between the parties, and also between the oscillators within each party. Enforcing this permutation symmetry, we take all $v^{ii} = a$ to be identical, as a consequence of which $V^{ij} (s) = v^{ij}$ is constant in $s$. Similarly, we take all $v^{ij} = b$ when $i \neq j$ but $ \hat{\mathcal{O}}_{ij} \in \bar{R}$ (irrelevant operators), and all $v^{ij} = c$ when $i \neq j$ and $ \hat{\mathcal{O}}_{ij} \in R$ (relevant operators). Therefore, by requiring total permutation symmetry, we have restricted the matrix of velocities to three independent parameters that determine the final unitary operator $\hat{U} (s)$. Of course, permutation symmetry between \emph{parties} as opposed to oscillators is not essential; for example, we could consider final states that are not symmetric in this way.  In Appendix~\ref{sec:ap2} we demonstrate how to compute  binding complexity for a less symmetric case and conjecture a solution for the completely general case.

Since the parameters $a$, $b$, and $c$ determine the operator $\hat{U} (s)$ that evolves from initial state to final state, we fix them by specifying these boundary conditions on the wavefunction. Namely, we take the initial wavefunction to be determined by the matrix $\Omega^{(i)} = \text{diag} (\omega_0, \omega_0, \ldots, \omega_0)$ and the final wavefunction to be determined by 
\begin{align}
\Omega^{(f)}_{ij} =\begin{cases} \omega , \qquad & i = j, \\ 
								\lambda_1, \qquad & \text{$i \neq j$ and $ \hat{\mathcal{O}}_{ij} \in \bar{R}$,} \\
								\lambda_2, \qquad &\text{$i \neq j$ and $ \hat{\mathcal{O}}_{ij} \in R$.} 								\end{cases} \label{eq:omegacomp}
\end{align}

Thus, the initial wavefunction is the product of Gaussians in every oscillator; it contains no entanglement. The final state wavefunction contains ``couplings'' $\omega$ of each oscillator to itself, couplings $\lambda_1$ between different oscillators in the same party, and couplings $\lambda_2$ between oscillators in different parties. The structure of the final wavefunction above is meant to be a permutation-symmetric toy model to mimic the structure of entanglement in a generic quantum field theory state, where if we partition our system into $m$ parties (i.e., either subregions or boundaries in the multiboundary case), then the state will have some internal entanglement within each party, in addition to entanglement between different parties. In $\Omega^{(f)}$, the couplings $\lambda_1$ create the internal entanglement between the oscillators inside each block/party, while the couplings $\lambda_2$ create entanglement between different blocks/parties. Although the wavefunction does not have the expected ``spatial locality'' of a quantum field theory state within each party, this locality can be added to the wavefunction by further acting on it with local unitary transformations which act only within each block; since such unitaries do not change the binding complexity, they will not affect our result below. For illustration, in the $N=3$ case, the matrix $\Omega^{(f)}$ takes the form
\begin{align}
\Omega^{(f)} = \left(\begin{array}{ccc|ccc|ccc|ccc} \omega & \lambda_1 & \lambda_1 & \lambda_2 & \lambda_2 & \lambda_2 & \quad & \quad & \quad & \lambda_2 &  \lambda_2& \lambda_2  \\  \lambda_1 & \omega & \lambda_1 & \lambda_2 & \lambda_2 & \lambda_2 & \quad & \ldots & \quad & \lambda_2 &  \lambda_2& \lambda_2 \\  \lambda_1 &  \lambda_1 & \omega & \lambda_2 & \lambda_2 & \lambda_2 & \quad & \quad & \quad & \lambda_2 &  \lambda_2& \lambda_2 \\ \hline   \lambda_2 & \lambda_2 & \lambda_2 & \omega & \lambda_1 & \lambda_1 & \quad & \quad & \quad & \lambda_2 &  \lambda_2& \lambda_2  \\  \lambda_2 & \lambda_2 & \lambda_2 &  \lambda_1 & \omega & \lambda_1 & \quad & \ldots & \quad & \lambda_2 &  \lambda_2& \lambda_2 \\  \lambda_2 & \lambda_2 & \lambda_2 & \lambda_1 &  \lambda_1 & \omega &  \quad & \quad & \quad & \lambda_2 &  \lambda_2& \lambda_2  \\ \hline &&&&&&&&&&& \\ &\vdots &&& \ddots &&& \ddots &&&\vdots & \\ &&&&&&&&&&& \\ \hline \lambda_2&\lambda_2&\lambda_2&\lambda_2&\lambda_2&\lambda_2&&&& \omega & \lambda_1 & \lambda_1 \\ \lambda_2&\lambda_2&\lambda_2&\lambda_2&\lambda_2 &\lambda_2&& \ldots &&\lambda_1 & \omega & \lambda_1 \\ \lambda_2&\lambda_2&\lambda_2&\lambda_2&\lambda_2&\lambda_2&&&& \lambda_1 & \lambda_1 & \omega   \end{array}  \right). \label{eq:omegablock}
\end{align}

Importantly, there are three independent couplings, matching the number of independent parameters of $V^{mn}$: $a$, $b$, and $c$. We determine the velocities $a$, $b$, and $c$ in terms of these couplings by examining how the matrix $\Omega$ flows under the infinitesimal action of the unitary $\hat{U} (s)$. Since $\hat{U}(s)$ does not take the wavefunction out of the set of Gaussian wavefunctions, we can label the state at an arbitrary time $s$ as
\begin{align}
|\Psi (s) \rangle = \hat{U} (s) |\Psi\rangle = \left(\frac{\det \Omega (s)}{\pi^n}\right)^{1/4} \int d\vec{\varphi}\: e^{-\frac12 \vec{\varphi}^T \Omega (s)  \vec{\varphi}} |\vec{\varphi} \rangle. \label{eq:gaussiantimevar}
\end{align}
Over an infinitesimal parameter length $ds$, the state changes according to
\begin{align}
\frac{d}{ds} |\Psi (s) \rangle = i \sum_{i,j} V^{ij}\hat{\cO}_{ij} |\Psi (s)\rangle. \label{eq:statechange}
\end{align}
This follows because (\ref{eq:statechange}) is a Schr\"{o}dinger equation, the solution of which for the operator $\hat{U}(s)$ is well-known to be the path-ordered exponential (\ref{eq:unitary}). 

Using the expression (\ref{eq:kops}) for the $\hat{\cO}_{ij}$ operators and $\hat{\pi}^i = -i\frac{\partial}{\partial \varphi^i}$ in the $|\varphi^i\rangle$ basis, the right-hand side becomes in this basis
\begin{align}
\langle \vec{\varphi} | i V^{ij}\hat{\cO}_{ij} |\Psi (s)\rangle = \left(-\frac12 \vec{\varphi} (2V\Omega ) \vec{\varphi} + \frac12 \text{Tr} (V) \right) \langle \vec{\varphi} | \Omega \rangle, \label{eq:opaction}
\end{align} 
where $V$ is the matrix of velocities $V^{ij}$. The symmetry of both $\Omega$ and $V$ has been used in deriving (\ref{eq:opaction}). The trace term can be absorbed into the wavefunction normalization, so the action of the $\hat{\cO}_{ij}$ operators induces the following flow of the matrix $\Omega$:
\begin{align}
\frac{d\Omega}{ds} = 2V\Omega. \label{eq:matrixflow}
\end{align}
Since $\Omega (s)$ has only three independent components $\omega(s)$, $\lambda_1 (s)$, and $\lambda_2 (s)$ by the ansatz (\ref{eq:omegacomp}), the  matrix equation (\ref{eq:matrixflow}) reduces to the three independent equations
\begin{align}
\frac{d\omega (s)}{ds} &= 2a\omega(s) + 2(N-1) b \lambda_1 (s) + 2N(m-1) c \lambda_2 (s) \label{eq:couplesystem1}\\
\frac{d\lambda_1 (s)}{ds} &= 2b\omega(s) + 2[a+(N-2)b] \lambda_1 (s) + 2N(m-1) c \lambda_2 (s) \\
\frac{d\lambda_2 (s)}{ds} &= 2c \omega (s) + 2c (N-1) \lambda_1 (s) + 2[a+(N-1)b + N(m-2)c] \lambda_2 (s). \label{eq:couplesystem3}
\end{align}
The coefficients above have been derived by expanding (\ref{eq:matrixflow}) and counting the number of terms of each type. Solving with the boundary conditions $\Omega^{(i)}$ and $\Omega^{(f)}$ specified earlier by taking $\omega(1) = \omega$, $\lambda_1 (1) = \lambda_1$, and $\lambda_2 (1) = \lambda_2$, we determine the constants $a$, $b$, and $c$. In terms of the three independent eigenvalues $(\lambda_+, \lambda_0, \lambda_-)$ of $\Omega$,
\begin{align}
\lambda_+ &= \omega + (N-1)\lambda_1 + N(m-1) \lambda_2 \label{eq:eigs1}\\
\lambda_0 &= \omega - \lambda_1 \\
\lambda_- &= \omega + (N-1)\lambda_1 - N\lambda_2, \label{eq:eigs3}
\end{align}
the constants are
\begin{align}
a &=\frac{1}{2mN} \ln \left(\frac{\lambda_+ \lambda_-^{m-1}}{\lambda_0^m} \right)  + \frac12 \ln \left( \frac{\lambda_0}{\omega_0}\right)\\
b &=  \frac{1}{2mN} \ln \left(\frac{\lambda_+ \lambda_-^{m-1}}{\lambda_0^m}\right)\\
c &= \frac{1}{2mN} \ln \left(\frac{\lambda_+}{\lambda_-}\right).
\end{align}
Plugging into (\ref{eq:complexity}) and counting the number of relevant operators $\hat{\mathcal{O}}_{ij} \in R$, the binding complexity of the general Gaussian wavefunction is therefore
\begin{align}
\mathcal{C}_b = N |c| \sqrt{m(m-1)}  =  \frac{1}{2}\sqrt{\frac{m-1}{m}} \left| \ln \left(\frac{\lambda_+}{\lambda_-}\right) \right| \, .
\end{align}
Unlike conventional circuit complexity \cite{Jefferson2017}, the binding complexity as computed here is finite in the $N \to \infty$ continuum limit of a large number of oscillators.

\begin{figure}[htbp!]
\begin{center}
\includegraphics[width=.8\textwidth]{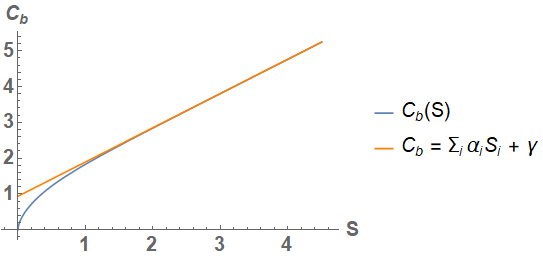}
\caption{When the entanglement entropy is large, the binding complexity varies linearly with the entropy up to exponentially small corrections.  Here the number of parties is $m=12$.  Other values of $m$ give similar results. \label{fig:linear}}
\end{center}
\end{figure}

We can also write the binding complexity in terms of the dimensionless, UV-finite parameter $\mu  = \frac{N\lambda_2}{\omega + (N-1) \lambda_1}$ as
\begin{align}
\mathcal{C}_b = \frac{1}{2}\sqrt{\frac{m-1}{m}}  \left| \ln \left(\frac{1+(m-1)\mu}{1-\mu}\right) \right|. \label{eq:complex2}
\end{align}
This parameterization is convenient because the entanglement entropy of a single party of oscillators relative to the rest is also controlled by $\mu$. Using the method of Srednicki \cite{Srednicki}, it is straightforward to compute that this entanglement entropy is
\begin{align}\label{eq:SPEntropy}
S = -\ln (1-\xi) - \frac{\xi}{1-\xi} \ln \xi, \qquad \xi = \frac{\beta'}{1+\sqrt{1-\beta^{'2}}}, \qquad \beta' =  \frac{(m-1) \mu^2}{2+2(m-2)\mu - (m-1) \mu^2}.
\end{align}
For $S$ to be finite, we must have $\frac{1}{1-m} < \mu < 1$; if we require $S$ to remain finite in the large $N$ limit, this similarly constrains $\frac{\lambda_2}{\lambda_1}$. At the points $\mu = 1$ and $\mu = \frac{1}{1-m}$, the entanglement entropy associated with a single party as well as the binding complexity both diverge. Expanding about either point, where the argument of the logarithm in (\ref{eq:complex2}) becomes large, as does the macroscopic entanglement entropy (i.e. we are at high temperature), we find that the binding complexity and  entanglement entropy are related as (also see Fig.~\ref{fig:linear}):
\begin{align}
\mathcal{C}_b =    \sum_{i=1}^m  \alpha_iS_i +\gamma+ \mathcal{O}(e^{-S}), \label{eq:scaling}
\end{align}
\beq
\alpha_i = \frac{1}{m} \sqrt{\frac{m-1}{m}}, \;\;\; \gamma = \sqrt{\frac{m-1}{m}} \left(\ln 2 - 1 + \frac12 \ln \frac{m^2}{m-1}\right),
\eeq
where $S_i$ refers to the entanglement entropy associated to the $i$th party. (For our symmetric wavefunctions all $S_i = S$ are equal). That is, the binding complexity scales linearly with the entanglement entropy, up to a constant term and  corrections exponentially small in the entropy.  As we will discuss below, this scaling of binding complexity with entropy resembles expectations from holographic duality.

That the binding complexity scales linearly with the entanglement entropy with both controlled by the same parameter $\mu$ is remarkable. Nevertheless, as discussed in the introduction, the single-party entanglement entropy may yield a misleading characterization of the robustness of entanglement in quantum states. To diagnose this robustness in the Gaussian states \eqref{eq:gaussian}, we use the Peres-Horodecki separability criterion as written by Simon \cite{PhysRevLett.84.2726}. This criterion is a necessary and sufficient condition for separability of a two-oscillator Gaussian state. Therefore, for the remainder of this section we work in the special case $N=1$, so that there are $m$ total oscillators with a single oscillator in each of the $m$ parties. We will check the separability of the reduced density matrix upon tracing out $m-2$ parties.

The Peres-Horodecki separability criterion is expressed in terms of the variance matrix $V_{ab} = \frac12 \langle  \Delta \hat{\xi}_a \Delta \hat{\xi}_b + \Delta \hat{\xi}_b \Delta \hat{\xi}_a \rangle$, with $\hat{\xi}_a = ( \hat{\phi}_1 , \hat{\pi}_1 , \hat{\phi}_2 , \hat{\pi}_2 )_a$ the phase-space coordinate operators of two oscillators and $\Delta \hat{\xi}_a = \hat{\xi}_a - \langle \hat{\xi_a} \rangle$. Writing $V$ in the block form $V = \begin{pmatrix} A & C \\ C^T & B \end{pmatrix}$ and defining the symplectic form $J = \begin{pmatrix} 0 & 1 \\ -1 & 0 \end{pmatrix}$, the density matrix $\rho$ is separable if and only if
\begin{align}
\mathcal{N}_g =  -\det A \det B - \left(\frac14 - |\det C| \right)^2 + \tr \left(AJCJBJC^TJ\right) + \frac14 (\det A + \det B) \leq 0 \label{eq:gaussnegativity}
\end{align}
We will call $\mathcal{N}_g$ the \emph{Gaussian negativity}. Taking $N=1$ and tracing out $m-2$ parties yields a state $\rho$ on two oscillators for which $\mathcal{N}_g$ evaluates to
\begin{align}
    \mathcal{N}_g = \frac{\mu^2}{4(1 - \mu)(1 + \mu (m-1))},
\end{align}
where $\mu = \frac{\lambda_2}{\omega}$ is the $N \to 1$ limit of the same parameter $\mu$ previously defined above \eqref{eq:complex2}. Recalling that $\frac{1}{1-m} < \mu < 1$ for the entropy and binding complexity to be finite, we see that this same condition leads to $\mathcal{N}_g > 0$. We conclude that the Gaussian states \eqref{eq:gaussian} for $N=1$ are robustly entangled like the W states. When $N>1$, the condition $\mathcal{N}_g \leq 0$ is no longer equivalent to separability \cite{PhysRevLett.86.3658}.\footnote{A criterion for the inseparability of Gaussian states with $N>1$ has been established \cite{PhysRevLett.95.230502} but requires an infinite series of inequalities to hold, which are difficult to check.} However, the similar structure of the wavefunction leads us to expect that the states will remain robustly entangled when $N > 1$.

\begin{figure}[h!]
\begin{center}
\includegraphics[width=0.8\textwidth]{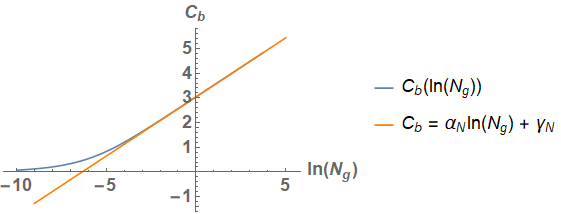}
\caption{The binding complexity varies linearly with the logarithm of the Gaussian negativity up to 
exponential corrections. Here the number of parties is again $m=12$ and other values of m give similar results.
\label{fig:linear2}}
\end{center}
\end{figure}

Since $\mathcal{N}_g$ is also controlled by the parameter $\mu$, we may again expand about the point where the binding complexity becomes large to find that the binding complexity scales linearly with the logarithm of the Gaussian negativity up to exponential corrections (see Fig.~\ref{fig:linear2}),
\begin{align}
    \mathcal{C}_b = \alpha_\mathcal{N} \ln \mathcal{N}_g + \gamma_\mathcal{N} + \mathcal{O}(\frac{1}{\mathcal{N}_g}),
\end{align}
with $\alpha_\mathcal{N} = \frac12 \sqrt{\frac{m-1}{m}}$, and $\gamma_\mathcal{N} = \sqrt{\frac{m-1}{m}} \log 2m$.

\begin{figure}[h!]
\begin{center}
\subfloat[\label{fig:3dcomplex}]{\includegraphics[width=.45\textwidth]{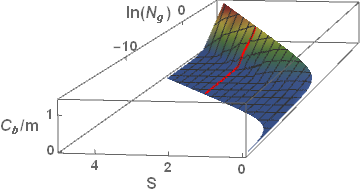}}
\subfloat[\label{fig:crosssec}]{\includegraphics[width=.45\textwidth]{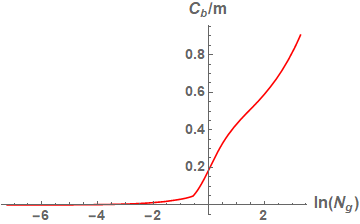}}
\caption{(a) The binding complexity per party $\mathcal{C}_b/m$ plotted versus $S$ and $\ln \mathcal{N}_g$ for a two-parameter family of states parameterized by $\mu$ and $m$. (b) A cross section of the left-hand side with fixed $S \approx 3.07$ chosen for plotting purposes, showing that $\mathcal{C}_b / m$ increases with $\ln \mathcal{N}_g$.
\label{fig:wormholesep}}
\end{center}
\end{figure}
Since the single-party entanglement entropy is controlled by the same parameter $\mu$ as the binding complexity $\mathcal{C}_b$ and the Gaussian negativity $\mathcal{N}_g$, it is not obvious if a large binding complexity ultimately stems from a robust entanglement structure rather than merely a large entanglement entropy. To address this question, Fig.~\ref{fig:wormholesep} shows that even at fixed entropy $S$, the binding complexity per party $\mathcal{C}_b /m$ increases with $\ln \mathcal{N}_g$. Consequently, binding complexity does diagnose robustness of entanglement.

\section{The interior volume of multiboundary wormholes} \label{sec:vol}

\begin{figure}[htbp!]
\begin{center}
\includegraphics[width=.5\textwidth]{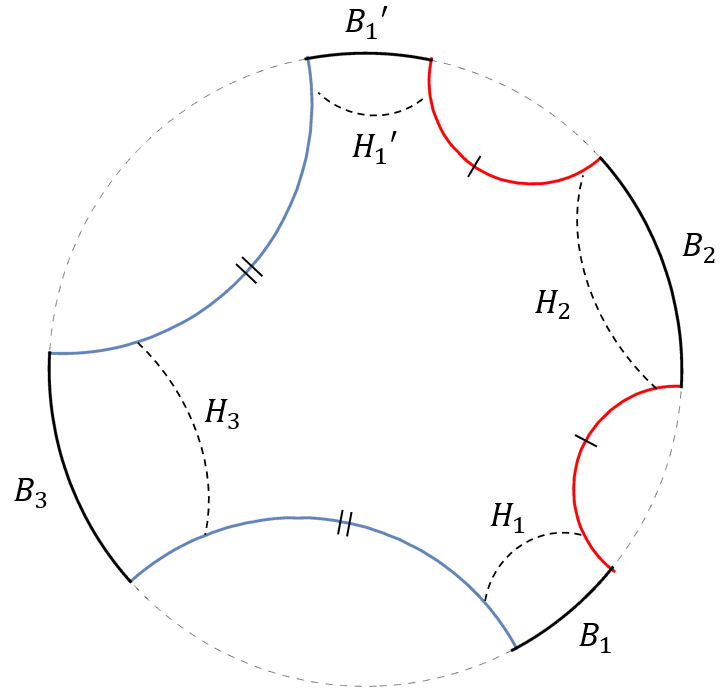}
\caption{Quotient construction of a three-boundary wormhole from vacuum $\text{AdS}_3$. Geodesics in blue and in red have been identified by the quotient, leading to boundary regions $B_1 \cup B_1'$, $B_2$, and $B_3$, with corresponding causal horizons $H_1 \cup H_1'$, $H_2$, and $H_3$ bounding an interior region. \label{fig:quotient1}}
\end{center}
\end{figure}

Multiboundary wormholes \cite{Brill1, Brill2, Brill3,Skenderis2011, Krasnov1, Krasnov2} are vacuum solutions of Einstein's equations in 2+1 dimensions that have multiple asymptotic regions (Fig.~\ref{fig:quotient1}).  Recently, properties of these geometries were used in \cite{Vijay2014, marolf, Fu2018} to investigate the entanglement structure and complexity of the boundary CFT state. Tensor network models for multiboundary wormholes were presented in \cite{ross}.

Like the two-sided BTZ black hole \cite{BTZ}, the multiboundary wormholes are constructed as quotients of $\text{AdS}_3$ space.   On the $t=0$ slice, $\text{AdS}_3$ is just hyperbolic space $\mathbb{H}^2$, which has an isometry group  $\text{PSL}(2,\mathbb{R})$.   The $t=0$ slice of the wormhole is obtained by quotienting this $\mathbb{H}^2$ by a discrete diagonal subgroup $\Gamma \subset \text{PSL}(2,\mathbb{R})$ with hyperbolic generators.
The action of  $\Gamma$ will identify pairs of boundary-anchored geodesics in $\mathbb{H}^2$, so $M = \mathbb{H}^2/\Gamma$ will be a Riemann surface with $m$ boundaries (each topologically $S^1$), where $m-1$ is the number of generators of $\Gamma$. Since any two disjoint boundary-anchored geodesics in $\mathbb{H}^2$ are joined by a unique minimal length geodesic, the endpoints of the latter join to form  causal horizons for the newly disjoint conformal boundary. The set of causal horizons bounds the interior of a wormhole that connects all the asymptotic regions together.  A holographic observer with access to observables on just a single boundary cannot access physics in the wormhole interior. It was shown in \cite{Vijay2014} that the CFT state dual to these wormholes contains multipartite entanglement between degrees of freedom localized on the different boundaries.

\begin{figure}[htbp!]
\begin{center}
\subfloat[\label{fig:tensor}]{\includegraphics[width=.4\textwidth]{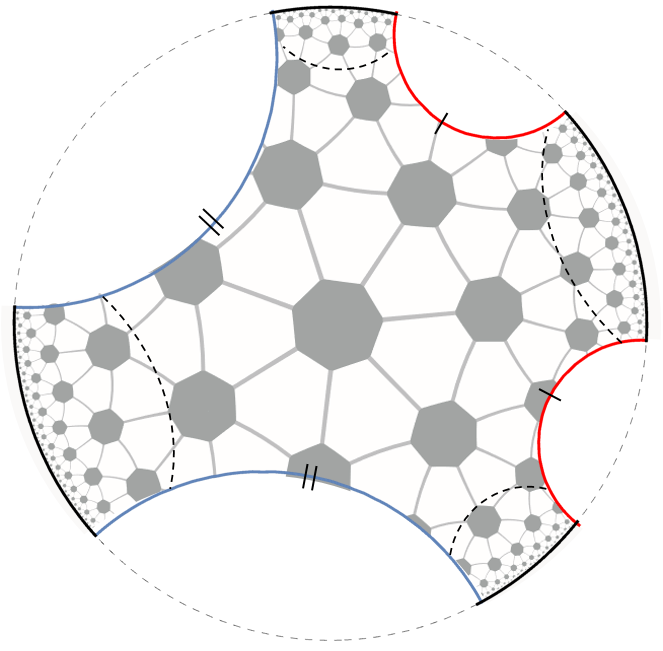}}\qquad\quad
\subfloat[\label{fig:distill}]{\includegraphics[width=.4\textwidth]{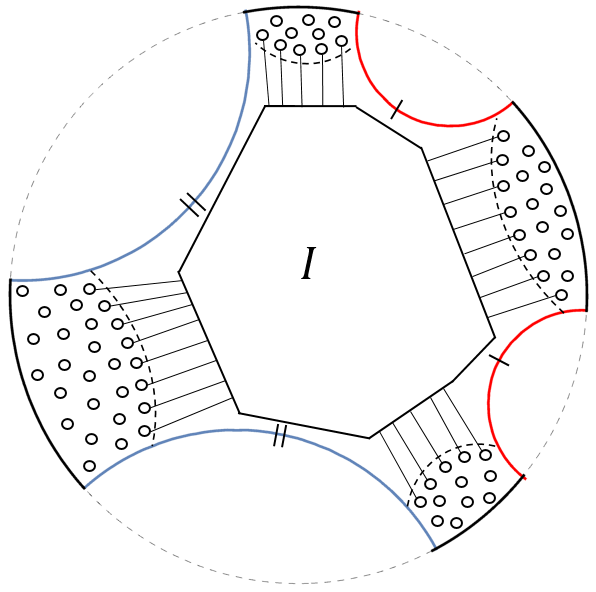}}
\caption{A schematic tensor network (a) preparing the boundary state dual to a three-boundary wormhole. In (b), the network has been distilled by local unitaries acting on each boundary, leaving sets of entangled bits lined up at each horizon as well as a multipartite residual region $I$. Stretching the horizons by $\ell_{\text{AdS}}$ captures nearby tensors contributing to the entanglement between boundaries.\label{fig:quotient23}}
\end{center}
\end{figure}

Following \cite{ross}, we can think of the complexity of the quantum state dual to a wormhole in holographic terms by imagining a tensor network that tiles the bulk Cauchy slice. Such a tensor network will prepare a state with the necessary pattern of entanglement (Fig.~\ref{fig:quotient23}a).  The complexity of the state is then proposed to be related to the number of gates in the tensor network \cite{1512.04993,XLQ2016,bartek, marolf, Fu2018}, an idea which correlates nicely with the proposal that complexity is holographically dual to the volume of spatial  slices \cite{1512.04993}.

In this tensor network construction of the boundary state dual to the wormhole, the tensors outside the horizons correspond to unitary operations acting {\it within} each boundary (Fig.~\ref{fig:quotient23}b). On the other hand, tensors enclosed within the wormhole interior may be thought of as corresponding to unitary quantum gates acting simultaneously on multiple boundaries. Thus, we might expect that \emph{binding complexity} corresponds to the interior volume of the wormhole. 

To make this comparison, we compute the interior volume of the wormhole. Since all of our calculations pertain to an equal-time slice of the  $2+1$-dimensional spacetime, the volume of the interior is really an area. It is easy to compute this area using the Gauss-Bonnet theorem in terms of the number of asymptotic boundary regions $m$, the genus of the interior $g$, and the geodesic curvature of each causal horizon. The interior $W_{g,m}$ is topologically a Riemann surface of genus $g$ with $m$ punctures, and the area of this surface, with the constant curvature metric inherited from $\mathbb{H}^2$, is given by
\begin{align}
\text{Area}(W_{g,m}) = 2\pi (2g+m-2) + \oint_{\partial W_{g,m}} k_g ds ,
\label{eq:interiorarea}
\end{align}
where the second term on the right hand side is the integral of the geodesic curvature on the boundary of the interior. 

We will set $g = 0$ for simplicity, i.e., the wormhole has a spherical internal topology.\footnote{In principle, we could extend our toy model construction to higher genus by considering states of more complicated entanglement structure (see Sec.~\ref{sec:pathint}).  For example, the ($m=4$, $g=1$) case might correspond to removing the $y_2$-$y_4$ and $y_1$-$y_3$ edges in Fig.~\ref{fig:branch3}.  The Euler-Arnold equation in the general form of this case becomes very difficult to solve, but such a calculation could serve as another check of our proposal that the (stretched) interior volume equals binding complexity.} If we choose the interior region to end strictly at the causal horizons (which are geodesic), then the geodesic curvature term vanishes.  In this case the area \eqref{eq:interiorarea} vanishes for the BTZ black hole (which has $m=2$).\footnote{This is because the causal horizons of the two asymptotic regions of the eternal BTZ black hole coincide on the $t=0$ surface at the bifurcation point of the horizon, so that, unlike the multiboundary case, the internal volume vanishes.}   Nevertheless, we know that there is bipartite entanglement between the two boundaries of BTZ, and there will be an associated binding complexity.  Thus the interior volume on the $t=0$ slice cannot be literally equal to complexity.  

In view of this, we are led to consider  ``stretched horizons", which are non-geodesic curves pushed slightly away from the true horizons in the full wormhole geometry toward the asymptotic boundaries (see \cite{lennystretch} and references therein). In the tensor network picture of complexity, we interpret this procedure as including tensors just outside the horizons which still contribute to the entanglement between multiple boundary CFTs, c.f. Fig.~\ref{fig:distill}. This interpretation is substantiated by \cite{ross}, which showed that for tensor network models built by quotienting the networks preparing vacuum $\text{AdS}_3$ states, it is possible for an  ``bipartite residual region" to remain after entanglement distillation in the $m=2$ case.   We are thinking of these residual tensors as living inside the stretched horizon. We will take the stretched horizon to be a surface of constant geodesic curvature $k_g$.

In sum, we obtain a contribution to the area that is proportional to the length of each stretched causal horizon
\begin{align}
\text{Area}(W_{0,m}) = 2\pi (m-2) + 4G\sum_{i=1}^m a_i S_i \, . \label{eq:interiorvol}
\end{align}
Here we used the fact that the horizon lengths are equal to $4G$ times the entropies of entanglement of the CFT on the $i^{\text{th}}$ boundary with all the other boundaries.\footnote{Note that assumes that we are in a region of the moduli space for the interior geometry of the wormhole where the entropy of each boundary is holographically given by the causal horizon separating it from the other asymptotic regions.  Remarkably, there are other regions of the moduli space where the entropy of boundary $i$ is actually give by the sum of areas of the causal horizons of all the {\it other} boundaries.  This surprising fact is explained in \cite{Vijay2014}.}
 The $\mathcal{O}(\ell^{-1}_{\text{AdS}})$ constants $a_i$ are given in terms of the horizon lengths by
\begin{align}
a_i L_i \equiv \oint_{\partial_i W_{0,m}} k_g ds ,
\end{align}
where $L_i$ is the horizon length and $\partial_i W_{0,m}$ is the $i^{\text{th}}$ boundary of the interior.

The formula (\ref{eq:interiorvol}) for the volume of the (stretched) wormhole is structurally similar to the formula (\ref{eq:scaling}) for binding complexity.  Both expressions have a constant piece, and a part that is linear in the entanglement entropies of each disconnected party.  Thus it is tempting to propose the correspondence\footnote{The volume here is being expressed in units of $\frac{1}{\ell_{\text{AdS}} G_N}$, as is usual in discussions of complexity.}
\begin{equation}
    \text{Binding Complexity} = \text{Volume of Stretched Wormhole Interior}
\end{equation}
In this correspondence, the factor $\frac{1}{m}\sqrt{\frac{m-1}{m}}$ in the binding complexity (\ref{eq:scaling}) plays the role of the coefficients $a_i$ in (\ref{eq:interiorvol}).    However, the constant term in the  binding complexity \eqref{eq:scaling} scales  as $\mathcal{O} (\ln m)$ and is nonzero for $m=2$, while the interior volume of the wormhole scales as $\mathcal{O}(m)$ and vanishes for $m=2$.
The origin of this discrepancy may lie in the simplicity of the toy model of Sec.~\ref{sec:mbycomplex} and might be resolved with an appropriate generalization of the framework for computing binding complexity of states in a nontrivial conformal field theory with a semiclassical bulk dual. However, it also might simply be that the toy model states whose complexity we considered were not structured in the same way as in holographic theories.
In Sec.~\ref{sec:pathint} we will provide evidence that the latter is indeed the case by using the Euclidean path integral to construct a natural class of states in our toy model whose binding complexity reproduces the form of the stretched wormhole volume.  Indeed in AdS$_3$/CFT$_2$ \cite{Vijay2014} precisely such a Euclidean procedure constructs the CFT states dual to the multiboundary wormhole.

\section{Euclidean path integrals}\label{sec:pathint}
In the previous section we argued that the binding complexity of Gaussian states that we calculated in Sec.~\ref{sec:mbycomplex} resembles the volume of the interior of multiboundary wormholes in AdS$_3$/CFT$_2$.   However, there was a discrepancy in the scaling of the complexity with the number of entangled parties $m$ which could arise if the permutation-symmetric states of Sec.~\ref{sec:mbycomplex} do not have the same entanglement structure as the states in AdS/CFT. In the AdS setting, the states dual to multiboundary wormholes can be constructed within the CFT by performing the Euclidean path integral on 2-manifolds with the topology of the bulk wormhole (i.e., the time-reflection symmetric Cauchy surface in the bulk) \cite{Vijay2014}.\footnote{In general, quantum field theory states on a $(d-1)$-dimensional Cauchy surface $\Sigma$ can be constructed by carrying out the Euclidean path integral on d-manifolds of different topologies and boundary $\Sigma$.}   To compare with the wormhole it would therefore be natural to compute the complexity of states in our toy model constructed in terms of similar Euclidean path integrals. In our case we have a collection of $n$ harmonic oscillators.  So, we should perform a path integral on a $(0+1)$-dimensional graph with $n$ external legs.  As will see, the binding complexity depends on the topology of the Euclidean graph.

A general 1D Euclidean path integral for a system of $n = Nm$ harmonic oscillators is computed on a graph $G$ consisting of a set of vertices $V_G$, $n$ of which are external, and a set of edges $E_G$ each of different lengths. Such a graph may contain internal vertices.  The value of the oscillator field at these vertices is a boundary condition which must be matched in the propagators at all incoming edges and integrated over.  Each edge $(v_1, v_2, \beta)$ of length $\beta$ between vertices $v_1, v_2$ at positions $x_1, x_2$ respectively in the graph corresponds to a factor of the propagator $K (x_1, x_2, \beta)$ in the integrand:
\begin{align}
K (x_1, x_2, \beta) = \langle x_2 | e^{-\beta H} | x_1\rangle = \int_{\phi(0) = x_1}^{\phi(\beta ) = x_2} [\mathcal{D}\phi] e^{-\int_0^\beta d\tau (\frac12 \dot{\phi}^2 + \frac12 M^2 \phi^2)} \, ,
\end{align}
where $\beta$ is the length of the edge in the graph and $M$ is oscillator mass.
The Euclidean propagator for the harmonic oscillator can be computed exactly; it is a Gaussian function known as the Mehler kernel:
\begin{align}\label{MK}
K(x_1, x_2, \beta) \propto \exp \left( -\frac{M((x_1^2+x_2^2)\cosh (M\beta ) - 2 x_1 x_2) }{2\sinh (M\beta)} \right). 
\end{align}

Let us label all external vertices by the vector $\vec{x}$ and internal vertices by the vector $\vec{y}$. The wavefunction of a state prepared by the Euclidean path integral on the graph $G$ is therefore
\begin{align}
\psi(\vec{x}) =  \int d\vec{y} \prod_{(v_1,v_2,\beta ) \in E_G} K(v_1,v_2,\beta ). \label{eq:pathintproc}
\end{align}
Since the propagator is Gaussian, the end result of the integrals over the internal vertices is also a Gaussian wavefunction, which can always be written
\begin{equation}
\psi(\vec{x}) = \mathcal{N} \exp \left( -\frac12 \vec{x}^T \Omega \vec{x} \right), \label{eq:graphgauss}
\end{equation}
where $\Omega$ is a real symmetric matrix and $\mathcal{N}$ is a normalization constant. Consequently, we may bring to bear the technology of Sec.~\ref{sec:mbycomplex} in computing the binding complexity.

\subsection{Permutation-symmetric graphs} \label{sec:branched}

We are interested in the complexity of states in which the different parties are multiparty entangled.  It is natural to imagine that such entanglement is produced in the Euclidean path integral if the graph is branched so as to connect between the parties.  In  Sec.~\ref{sec:mbycomplex} we considered states \eqref{eq:omegacomp} in which the oscillators within parties were entangled with one strength, while the parties as a whole were entangled block-wise with other parties and with a different strength.  We will first see how to construct such permutation-symmetric states through a Euclidean path integral.

\begin{figure}[h!]
\begin{center}
\subfloat[\label{fig:branch1}]{\includegraphics[width=.3\textwidth]{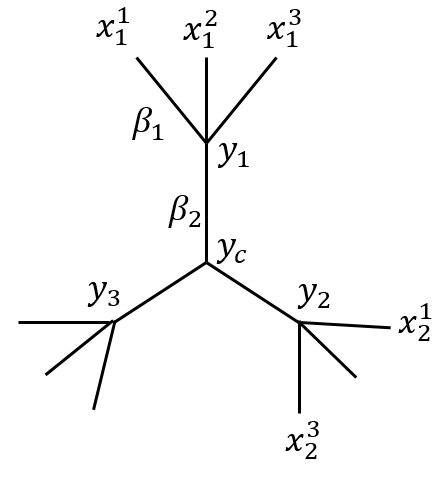}}
\subfloat[\label{fig:branch2}]{\includegraphics[width=.3\textwidth]{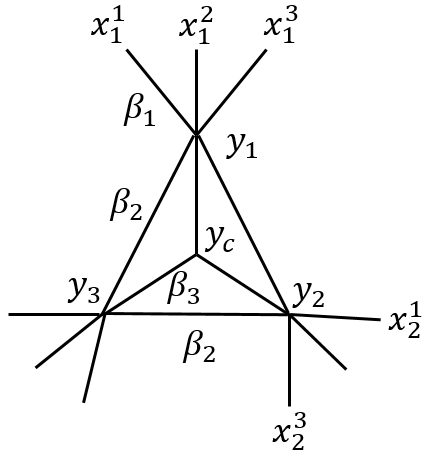}}
\subfloat[\label{fig:branch3}]{\includegraphics[width=.35\textwidth]{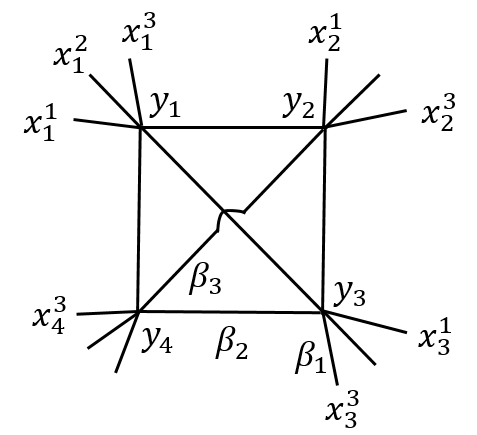}}
\caption{Three similar branched graphs with different internal topology. On the left, (a) displays the simplest case for $m=3$ and $N=3$ where the lines meet at a central vertex $y_c$. In the middle and on the right, (b) and (c) demonstrate a higher degree of internal connectedness. The internal lines of (b) form a complete graph including a central vertex $y_c$, while (c) is identical except for the removal of the central vertex. We have taken $m=3$ for (b), $m=4$ for (c), and $N=3$ for both.  Euclidean path integrals on all these graphs produce states that are permutation-symmetric between the parties.
\label{fig:branched}}
\end{center}
\end{figure}

In the Euclidean path integral, oscillators become entangled if their propagators meet at a vertex where a shared boundary condition is integrated over.  This suggests that to construct the states in the previous section we need a graph with $m$ groups of $N$ external lines that each meet at a vertex to create the internal entanglement within parties.  These vertices can then be connected by further propagators to create entanglement between the parties.  Three such graphs are shown in Fig.~\ref{fig:branched}.    We label the vertices at the end of the external lines as $x_j^i$ for the $i$th oscillator in the $j$th party.   The internal vertices can have any number of lines ending on them -- the internal structure of the graph can be completely arbitrary up to the permutation symmetry of the state we are trying to construct.  In analogy with the holographic setting, we might refer to the internal part of the graphs in Fig.~\ref{fig:branched} as a ``wormhole'' connecting the exterior legs.

First consider the simplest graph Fig.~\ref{fig:branch1}.  The internal vertices on the $i$th branch are labeled $y_i$, and the central vertex is labeled $y_c$.  We integrate over the boundary condition of the field at each vertex to perform the path integral.  The lengths of the edges are moduli of the graph, and the wavefunction generated by the path integral is a function of these parameters.   Permutation symmetry of the states \eqref{eq:omegacomp} dictates that the external lines have the same length ($\beta_1$) and the internal lines have the same length ($\beta_2$).   Similarly, Figs.~\ref{fig:branch2},~\ref{fig:branch3} have three moduli.

Performing the path integral on the family of graphs of Fig.~\ref{fig:branch1} according to the procedure of (\ref{eq:pathintproc}), one obtains  a Gaussian state (\ref{eq:graphgauss}) in the permutation-symmetric form  (\ref{eq:omegacomp}) with parameters $\omega,\lambda_1,\lambda_2$ where $\omega$ and $\lambda_1$ quantify entanglement within a party and $\lambda_2$ quantifies entanglement between parties.  We find that (see Appendix~\ref{sec:ap1} for details)
\begin{align}
\omega &= \frac{\omega_N}{\omega_D} \label{eq:omegcouple}\\
\lambda_1 &= \frac{M \text{csch}^2(M \beta_1) \coth (M \beta_2) (m N \coth (M  \beta_1)+m \coth (M \beta_2)-2 (m-1) \text{csch}(2 M \beta_2))}{m (N \coth (M  \beta_1)+\coth (M \beta_2)) (N \coth (M  \beta_1) \coth (M \beta_2)+1)}\\
\lambda_2 &= \frac{M \text{csch}^2(M \beta_1) \text{csch}^2(M \beta_2)}{m (N \coth (M \beta_1)+\coth (M \beta_2)) (N \coth (M \beta_1) \coth (M \beta_2)+1)} \, , \label{eq:massivecouple}
\end{align}
with
\begin{align}
\omega_N &= M \biggl(m N^2 \coth ^3(M \beta_1) \coth (M \beta_2)+m N \coth ^2(M \beta_1) \left(\coth ^2(M \beta_2)+1\right) \nonumber\\
&\qquad\quad +m \coth (M \beta_1) \coth (M \beta_2) \left(1-N \text{csch}^2(M \beta_1)\right) \nonumber\\
&\qquad\quad -\frac{1}{2} \text{csch}^2(M \beta_1) \text{csch}^2(M \beta_2) (m \cosh (2 M \beta_2)-m+2)\biggr) \\
\omega_D &= m (N \coth (M \beta_1)+\coth (M \beta_2)) (N \coth (M \beta_1) \coth (M \beta_2)+1).
\end{align}
Since this is a permutation-symmetric Gaussian state, the binding complexity is given by \eqref{eq:scaling} and the entanglement entropy of a single party is given by \eqref{eq:SPEntropy}.  Both quantities vanish in the limit $N \to \infty$ with $\beta_1, \beta_2, M$ fixed since $\lambda_2$ (which quantifies entanglement between parties) scales as $1/N^2$ at large $N$.   This disentangling at large $N$ can be understood as a manifestation of the principle of entanglement monogamy: when the number of oscillators within a party  grows large, most  oscillators are entangled within their party rather than with other parties. We can compensate by taking a kind of 't Hooft limit in which $\beta_1 N$ is held fixed as $N \to \infty$, in which case both the binding complexity and entanglement entropy will be finite and nonzero since $\lambda_2$ approaches a finite value in the large $N$ limit. The latter scaling limit can also be thought of as a rescaling of the couplings with the lattice scale so that the couplings remain finite in the continuum for a lattice quantization of scalar field theory. 

\begin{figure}[htbp!]
\begin{center}
\includegraphics[width=.6\textwidth]{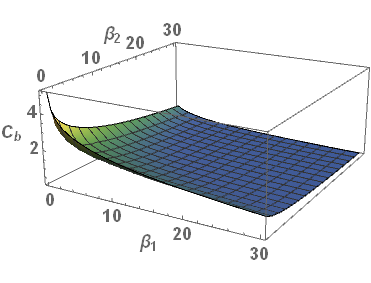}
\caption{Binding complexity for states constructed by the Euclidean path integral on the graph Fig.~\ref{fig:branch1} as function of the moduli.  Here, for illustration, we take $m=12$, $N=20$, and $M=.01$. \label{fig:modulidep}}
\end{center}
\end{figure}

Fig.~\ref{fig:modulidep} shows the moduli dependence of the binding complexity for the graph Fig.~\ref{fig:branch1} as computed in \eqref{eq:omegcouple}-\eqref{eq:massivecouple}. 
The complexity increases as $\beta_1,\beta_2$ become small.  This is because as $\beta \to 0$ the propagator in \eqref{MK} becomes the identity, thus more closely coupling the values of the fields at either end of a line in the graph.  In the other limit, as $\beta \to \infty$, the propagator projects onto the ground state, essentially decoupling the external oscillators from the internal structure of the graph.    Finally, consider wavefunctions associated with the graphs Fig.~\ref{fig:branch2} and Fig.~\ref{fig:branch3}.  Because all the integrals are Gaussian, we will again get Gaussian wavefunctions and because the graphs are permutation-symmetric, the wavefunctions will be as well. Of course, the coefficients in the wavefunctions will contain different functions of the moduli in each case because the detailed integrals are different.  However, all of these wavefunctions are necessarily of the form \eqref{eq:omegacomp}, and therefore the constant term in the binding complexity will scale as the logarithm of the number of entangled parties, unlike the linear scaling with parties of the interior volume of multiboundary wormholes. 

\subsection{Bipartite entanglement graphs} \label{sec:local}

We would like to find graphs that generate states with complexity-entropy scaling relations that match the holographic form. First note that the scaling relation (\ref{eq:scaling}) between complexity and entropy holds in the large $\beta$ limit in which the entropy associated with any single party is large. It was shown in \cite{marolf} in the holographic setting that in this regime, the entanglement structure of the multiboundary wormhole is dominated by bipartite entanglement between boundaries.\footnote{This was justified by computations of the mutual information both from the CFT state dual to the wormhole and holographically using the Ryu-Takayanagi formula. The tensor network models for multiboundary wormholes considered in \cite{ross} corroborate the dominance of bipartite entanglement in the large $\beta$ limit.} Consequently, to better match the holographic expectations, we seek graphs on which the path integral will produce strongly bipartite entanglement. There is an independent reason to be interested in such graphs: in the ``bit thread" interpretation of holographic entanglement entropy \cite{headrick, hubeny} one expects the correlations between independent tensor factors of a CFT to be dominated by bipartite entanglement (i.e., between the two qubits connected by a bit thread).  In our setup, the mixing of terms in the wavefunction is dictated by topological connectedness in the graph on which we perform the path integral. Therefore, we engineer multipartite entangled states with locally bipartite entanglement structure by using graphs which factorize so that a given connected component of the graph connects only two parties.

\begin{figure}[htbp!]
\begin{center}
\subfloat[\label{fig:bipartite1}]{\includegraphics[width=.45\textwidth, valign=c]{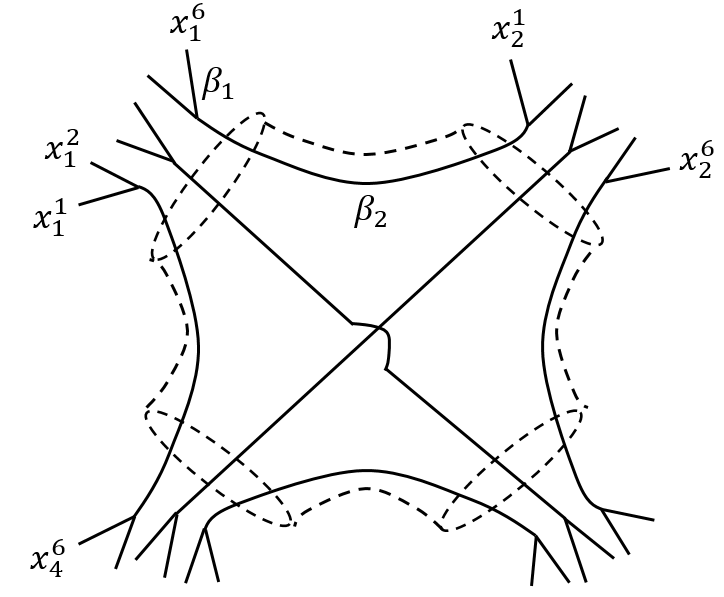}}
\subfloat[\label{fig:bipartite2}]{\includegraphics[width=.45\textwidth, valign=c]{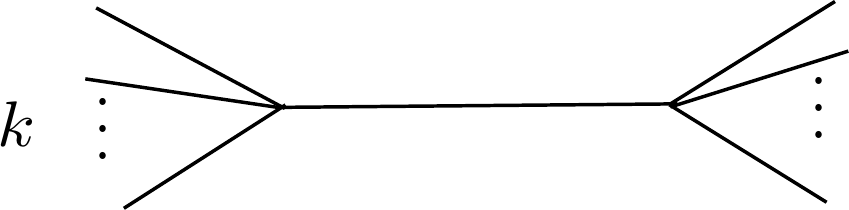}\vphantom{\includegraphics[width=0.45\textwidth,valign=c]{mbdybipartite.png}}}
\caption{(a) The $m=4$, $N=6$, $k=2$ bipartite entanglement graph. Note that there is no central vertex in the interior; the graphs overlap each other but are not connected. (b) The graph corresponding to $\psi_{\mathrm{branch},k}$, with $k$ oscillators in each of the two parties. \label{fig:localent}}
\end{center}
\end{figure}

In Fig.~\ref{fig:bipartite1}, we display such a ``bipartite entanglement graph", in which the oscillators in each party have been partitioned into groups that are only entangled with oscillators in one other party. The overall graph factorizes into a collection of the two-party permutation-symmetric graphs of Sec.~\ref{sec:branched}. Let $N=(m-1)k$ be the number of oscillators in each party, where $m$ is the total number of parties and $k$ is the number of oscillators per grouping, so that each of the $k$ groups connects to a different one of the other $m-1$ parties (see Fig.~\ref{fig:localent} for details). We again choose $\beta_2$ to be the length of internal lines and $\beta_1$ to be the length of external lines.  As drawn in Fig.~\ref{fig:bipartite1} it appears that only part of each party is connected to part of another party. However, as before, one may always mix the oscillators in a single party via local unitaries which will not affect the binding complexity or the entanglement entropy associated with that party. Therefore, we may think of Fig.~\ref{fig:bipartite1} as encoding locally bipartite entanglement between parties without restricting the entanglement to reside in some subsystem of each party. In other words, in Fig.~\ref{fig:bipartite1} we have used local unitary transformations to ``diagonalize'' the entanglement structure in each party.

Since the manifold on which we are performing the Euclidean path integral is topologically disconnected, the path integral factorizes over the connected components, as does the resulting wavefunction. Consequently, $\psi(\vec{x})$ is the product of ${m \choose 2} = \frac{m(m-1)}{2}$ permutation-symmetric wavefunctions. Let $\psi_{\text{branch},k}$ be the wavefunction of the graph in Fig.~\ref{fig:bipartite2}, which has $k$ oscillators in each party. This is a permutation-symmetric graph as described in Sec.~\ref{sec:branched}, so $\psi_{\text{branch},k}$ is a two-party permutation-symmetric wavefunction. Then the wavefunction of the full bipartite entanglement graph can be explicitly written as
\begin{align}
\psi (\vec{x}_1, \ldots, \vec{x}_m) &= \psi_{\text{branch},k}(x_1^{(m-2)k+1}, \ldots, x_1^{(m-1)k}, x_2^1, \ldots, x_2^k)\times \ldots \nonumber \\
&\quad\times \psi_{\text{branch},k}(x_m^{(m-2)k+1}, \ldots, x_m^{(m-1)k}, x_1^1, \ldots, x_1^k), \label{eq:wffactorized}
\end{align}
where the product includes $m(m-1)/2$ such terms corresponding to the bipartite connection between each pair of parties. The total wavefunction is still Gaussian and takes the form of (\ref{eq:graphgauss}), but $\Omega$ is no longer permutation symmetric within each party. Since we have the freedom to relabel oscillators so that topologically connected vertices are ordered adjacently in the matrix, $\Omega$ takes a block-diagonal form, consisting of $m(m-1)/2$ identical permutation-symmetric subblocks each of size $2k \times 2k$. Each subblock is of the form (\ref{eq:omegacomp}) with the couplings $\omega$, $\lambda_1$, and $\lambda_2$ given by (\ref{eq:omegcouple}) - (\ref{eq:massivecouple}) with the replacement $m \to 2$ and $N \to k$.  The structure of the matrix $\Omega$ in the special case $m=3$, $N=4$, $k=2$ is shown below, where the solid lines demarcate parameters corresponding to the same party and dashed lines demarcate parameters corresponding to topologically connected oscillators:
\begin{align}
\Omega= \left(\begin{array}{cc;{2pt/2pt}cc|cc;{2pt/2pt}cc|cc;{2pt/2pt}cc}
 \omega & \lambda_1 & 0 & 0    & 0 & 0 & 0 & 0        & 0 & 0 &  \lambda_2 & \lambda_2 \\ 
 \lambda_1 & \omega &  0 & 0    & 0 & 0 & 0 & 0     & 0 & 0 &  \lambda_2 & \lambda_2  \\ \hdashline[2pt/2pt] 
0&  0 & \omega & \lambda_1 & \lambda_2 & \lambda_2 & 0 & 0     & 0  & 0 & 0 & 0 \\ 
0 & 0 & \lambda_1 & \omega & \lambda_2 & \lambda_2 & 0 & 0 & 0 & 0 & 0 & 0  \\ \hline  
 0&  0 & \lambda_2 & \lambda_2 & \omega & \lambda_1 & 0 & 0  & 0 & 0 & 0 & 0  \\
 0 & 0 & \lambda_2 & \lambda_2 & \lambda_1 & \omega & 0 & 0  & 0 & 0 & 0 &  0 \\ \hdashline[2pt/2pt] 
 0 & 0 &0 &0   &  0 & 0 & \omega & \lambda_1 & \lambda_2 & \lambda_2 & 0 & 0  \\
 0 & 0 & 0 & 0    &  0 & 0 & \lambda_1 & \omega & \lambda_2 & \lambda_2 & 0 & 0 \\ \hline 
0&0&0&0&0&0&\lambda_2 & \lambda_2 & \omega & \lambda_1 &0&0\\ 
0&0&0&0&0&0&\lambda_2 & \lambda_2 & \lambda_1 & \omega  &0&0 \\ \hdashline[2pt/2pt] 
\lambda_2 & \lambda_2 &0&0&0&0&0 & 0 & 0 & 0 & \omega & \lambda_1 \\ 
\lambda_2 & \lambda_2 &0&0&0&0&0 &0& 0 & 0 & \lambda_1 & \omega  
\end{array}  \right). \label{eq:localomega}
\end{align}
We have not yet relabeled oscillators above to bring $\Omega$ into block diagonal form, so that the grouping of oscillators in each party is clearer.

Although $\Omega$ is not permutation-symmetric, each of its subblocks \emph{is} permutation-symmetric. Consequently, the entanglement entropy associated with a single party is
\begin{align}
S = (m-1)S_{\text{branch},k}, \label{eq:bitee}
\end{align}
where $S_{\text{branch},k}$ refers to the entanglement entropy associated with a single party of the wavefunction $\psi_{\text{branch},k}$. Equation (\ref{eq:bitee}) follows automatically from the factorized form of the graph as shown in Fig.~\ref{fig:bipartite1}: the wavefunction splits over each component in the graph, so the total entanglement entropy is the sum of the entropies of each component.\footnote{This follows from the property $S(\rho_A \otimes \rho_B) = S(\rho_A) + S(\rho_B)$.} In other words, the entanglement entropy associated to a single party essentially counts the minimal number ($m-1$) of edges which are ``cut" in separating the oscillators in that party from the rest. This continues to hold for the entropy associated with other partitions: the prefactor $m-1$ in (\ref{eq:bitee}) changes to the minimal number of edges cut in separating those parties from the rest. It is tempting to compare this result to bit threads and to the tensor network picture of holographic entanglement entropy, in that the entropy associated to a given party is directly proportional to the number of ``threads"  leaving that party.  This counting property of entropy is thought to underlie the Ryu-Takayanagi formula for holographic entanglement entropy.

Upon tracing out $m-2$ parties, the reduced density matrix $\rho$ associated with two parties of a bipartite entanglement graph has a robust W-like entanglement structure. From the product structure of the wavefunction \eqref{eq:wffactorized}, it follows that $\rho$ takes the schematic form $\rho = \rho_1^{\text{mixed}} \otimes \rho_{12} \otimes \rho_2^{\text{mixed}}$, where the subscripts refer to the first and second party. Here $\rho_{12}$ is a pure state corresponding to the two-party permutation-symmetric graph that connects a single group of oscillators in each of the two parties, while $\rho_1^{\text{mixed}}$ refers to the complicated mixed state of the remaining oscillators in the first party and similarly for $1 \leftrightarrow 2$. In Sec.~\ref{sec:mbycomplex} we argued that a permutation-symmetric state like $\rho_{12}$ has a robust entanglement structure, so $\rho$ will demonstrate this structure as well. In Fig.~\ref{fig:bipartite1} and in \eqref{eq:wffactorized}, we have picked an adapted basis that has separated the oscillators in such a way that upon doing partial traces, the degrees of freedom that remain entangled are distinct from the degrees of freedom that are in a mixed state.  In general, we can act with local unitary transformations so that all the degrees of freedom retain both entanglement and mixedness.

The binding complexity of these graphs is $\sqrt{{m \choose 2}}$ times the binding complexity of $\psi_{\text{branch},k}$ as computed by \eqref{eq:complex2}, giving
\begin{align}\label{eq:complex3}
    \mathcal{C}_b = \frac14\sqrt{m(m-1)}\left|\ln \left(\frac{1+\mu}{1-\mu} \right) \right| \, .
\end{align}
Here $\mu = \frac{k\lambda_2}{\omega+(k-1)\lambda_1}$ as is appropriate for $\psi_{\text{branch},k}$. Equation~\eqref{eq:complex3} follows from the factorized nature of the wavefunction, since the minimal circuit preparing the final state splits over each of the ${m \choose 2}$ components in the graph, as can be checked by explicitly solving the Euler-Arnold equation.  This splitting leads to an overall factor of $\sqrt{{m \choose 2}}$ from the sum over different factors \emph{inside} the square root in the equation \eqref{eq:complexity} for complexity. 

The complexity-entropy scaling relation that follows from \eqref{eq:bitee} and \eqref{eq:complex3} is
\begin{align}
\mathcal{C}_b = \frac12 \sqrt{m(m-1)}(2\ln 2 - 1) + \sum_{i=1}^m \frac{1}{2m} \sqrt{\frac{m}{m-1}} S.
\end{align}
Comparing to (\ref{eq:interiorvol}), one sees that the constant term now scales with the number of boundaries in the same  $\mathcal{O}(m)$ fashion as the holographic expectation, at least in the large $m$ limit, adding support for the idea that Binding Complexity = Wormhole Volume.

\section{Complexity for coherent states in perturbation theory} \label{sec:pttheory}

The Nielsen formalism also allows us to compute how much the complexity of a state changes when it is perturbed. For example, suppose we want to compute the Nielsen complexity of a state of the form
\beq
| \psi \rangle  = e^{it\sum_I h^I \cO_I} | \psi_0\rangle,
\eeq
relative to the base state $\psi_0$, where $t$  will be treated as a small parameter in which we do perturbation theory. In other words, we are interested in studying the complexity of the unitary operator $U =e^{it\sum_I h^I \cO_I}$ perturbatively in $t$. This situation can arise in several contexts; for example if we treat $t$ as time and $H = \sum_I h^I \cO_I$ as a Hamiltonian, then we obtain the small-time behavior of the complexity of time evolution. Alternatively, we may treat $U$ as creating a \emph{coherent state} on top of some base state $\psi_0$, and $t$ may be a small parameter which controls the size of the background deformation, as in the next subsection.

As before, we will take the operators $\cO_I$ to form the Lie algebra
\beq
\left[\cO_I, \cO_J \right] = i\sum_K {f_{IJ}}^K\cO_K.
\eeq
As discussed previously, we need to define a positive-definite, bilinear form $G_{IJ}$ on this algebra, which fixes the complexity of individual gates. We will be interested in the case of binding complexity, where operators which act within individual factors will have small cost factors, while operators which act across multiple factors will have $O(1)$ cost factors. From $G_{IJ}$ we can then define a right-invariant metric on the entire group manifold by pulling back this bilinear form from the identity. A geodesic takes the general form
\beq
|\psi(s) \rangle = U(s) | \psi_0\rangle, \;\;\; U(s) = \mathcal{P} \exp\left(i\int_0^s ds' \sum_I v^I(s')\cO_I\right),
\eeq
where $v$ is the local velocity and $\mathcal{P}$ is path-ordering. The geodesic equation in terms of the velocity is given by the Euler-Arnold equation 
\beq
\sum_J \cI^I_{J} \frac{dv^J}{ds} - \sum_{K,L,M} {f_{KL}}^Iv^K \cI^L_Mv^M = 0,
\eeq
The boundary conditions are
\beq \label{bc}
U(0) = 1, \;\;\; U(1) = \exp\left(it \sum_I h^I\cO_I\right),
\eeq
where $\lambda$ is a small parameter. For states of this form, we can solve the equations in perturbation theory with respect to $\lambda$. So let us take
\beq
v^I(s) = t v^I_{(1)}(s) + t^2 v^I_{(2)}(s) + t^3 v^I_{(3)}(s) + \cdots.
\eeq

\noindent \textbf{First order in $\mathbf{t}$}: At leading order, the equation can be solved trivially:
\beq
\frac{dv^I_{(1)}}{ds} = 0 \;\; \Rightarrow \;\; v^I_{(1)}(s) = v^I_{(1)}(0),
\eeq
Therefore, the unitary $U(1)$ is given by
\beq
U(1) = 1 + i t \sum_I v^{I}_{(1)}\cO_I+\cdots.
\eeq
Comparing this with \eqref{bc} at first order, we deduce that
\beq 
v^{I}_{(1)} = h^I.
\eeq
We can now compute the binding complexity of this state as the geodesic distance:
\beq
\mathcal{C}_b := \int_0^1 ds\,\sqrt{\sum_{I,J} G_{IJ}v^I(s) v^J(s)} = t || \mathbf{H} ||+O(t^2),
\eeq
where $|| \mathbf{H} || = \sqrt{\sum_{I,J} G_{IJ}h^I h^J}$ is the norm of the operator $\mathbf{H} = \sum_I h^I\cO_I$ with respect to the chosen complexity metric $G$. 

\noindent\textbf{Second order in $\mathbf{t}$}: At the next order in $t$, we find the solution
\beqn
v^I_{(2)}(s) &=& v^I_{(2)}(0) + s \sum_{J,K,L,M} {f_{KL}}^J(\cI^{-1})^I_Jh^K\cI^L_Mh^M \nonumber\\
&=& v^I_{(2)}(0) + s\sum_{K,M} \,{c_{KM}}^Ih^Kh^M,
\eeqn
where we have defined 
\beq
{c_{KM}}^I = \sum_{J,L} {f_{KL}}^J(\cI^{-1})^I_J \cI^L_M.
\eeq 
So, now the unitary becomes
\beqn
U(1) &=& \mathcal{P}\exp\left(i \int_0^1 ds\,\sum_I \left[t h^I + t^2\left(v^I_{(2)}(0) + s \sum_{K,L} {c_{KL}}^Ih^Kh^L\right)\right]\cO_I+\cdots\right)\nonumber\\
&=& 1 +i \sum_I\left[t h^I + t^2\left(v^I_{(2)}(0) +\frac{1}{2} \sum_{K,L} {c_{KL}}^Ih^Kh^L\right)\right]\cO_I \\
&&\quad - \frac{t^2}{2} \sum_{I,J} h^Ih^J\cO_I\cO_J+\cdots.
\eeqn
Once again, comparing with equation \eqref{bc}, we find 
\beq
v^I_{(2)}(0) =-\sum_{K,L} {c_{KL}}^Ih^Kh^L,
\eeq
and therefore to this order the velocity is then given by 
\beq
v^I(s) =t h^I- \frac{1}{2}t^2\left(1-2s\right) \sum_{K,L} {c_{KL}}^Ih^Kh^L+ \cdots.
\eeq
We can now use this result to compute the $O(t^2)$ correction to the complexity, and we find that the $O(t^2)$ contribution vanishes after performing the $s$-integral. Therefore, we obtain
\beq
\mathcal{C}_b =t || \mathbf{H} ||+O(t^3).
\eeq
We can proceed in a similar fashion to obtain higher order corrections, for instance, the $O(t^3)$ correction is shown in Appendix \ref{app3}. We see that for small $t$ the binding complexity of the unitary $U = e^{i t \sum_Ih^I\cO_I}$ increases linearly in $t$, with the proportionality constant being the norm of the Hamiltonian $H = \sum_I h^I\cO_I$ in the multipartite sector, that is, only the relevant operators which act simultaneously on multiple factors are included in the norm. 

\subsection{Double-trace deformations: towards creating wormholes}

We can now how ask how the binding complexity changes if we perturb a state by acting with an operator that locally couples degrees of freedom in two distinct parties.  In the holographic context this sort of ``double-trace deformation'' was shown in \cite{Gao2017} to create or expand a wormhole in the geometric description of disconnected but entangled CFTs.  Once again we consider the toy model with $n$ free, decoupled harmonic oscillators $\phi_i$, with the Hamiltonian
\beq
H_0 = \frac{1}{2} \sum_i\left(\hat{\pi}_i^2 + \hat{\phi}_i^2\right).
\eeq
We can straightforwardly diagonalize $H_0$ by introducing the creation and annihilation operators $a_i = \frac{1}{\sqrt{2}}\left(\hat{\phi}_i + i\hat{\pi}_i\right)$ and $a_i^{\dagger} = \frac{1}{\sqrt{2}}\left(\hat{\phi}_i - i\hat{\pi}_i\right)$, in terms of which we obtain $H_0 =\sum_i  \left(a^{\dagger}_i a_i + \frac{1}{2}\right)$. The vacuum state for this Hamiltonian, which satisfies
\beq
a_i | \psi_0 \rangle = 0,
\eeq
is a completely decoupled product state, and as such it will have no binding complexity. We now deform the Hamiltonian by a small bilinear coupling
\beq
H = H_0 + H_{\text{int}},
\eeq
with
\beq
H_{\text{int}} = \frac{g}{2} \sum_{i,j} \hat{\phi}_i C_{ij}\hat{\phi}_j.
\eeq
The  coupling clearly introduces some entanglement and binding complexity in the new vacuum; our aim here is to compute this binding complexity perturbatively in $g$. In order to diagonalize the new Hamiltonian $H$, let us introduce the orthogonal matrix $V_{ij}$ which diagonalizes $M_{ij} = \delta_{ij} + gC_{ij}$:
\beq
M = V^T\cdot D \cdot V,\;\;\; D = \mathrm{diag}\left(\omega_1^2, \omega_2^2 ,\cdots\right).
\eeq
Here $\omega_i^2$ are the eigenvalues of $M$.  Then, we define the new operators
\beq
\Phi_i = \sum_j V_{ij}\hat{\phi}_j,\;\;\; \Pi_i = \sum_j V_{ij}\hat{\pi}_j,
\eeq
which also satisfy the appropriate bosonic commutation relations. In terms of these new variables the full Hamiltonian becomes
\beq
H = \frac{1}{2} \sum_i \Pi_i^2 + \frac{1}{2}\sum_i \omega_i^2 \Phi_i^2.
\eeq
Now diagonalize this Hamiltonian by introducing the new creation and annihilation operators 
\beq
A_i = \frac{1}{\sqrt{2}}\left(\sqrt{\omega_i}\Phi_i + \frac{i}{\sqrt{\omega_i}}\Pi_i\right), \;\;A^{\dagger}_i = \frac{1}{\sqrt{2}}\left(\sqrt{\omega_i}\Phi_i - \frac{i}{\sqrt{\omega_i}}\Pi_i\right).
\eeq
We can express these new creation and annihilation operators in terms of the old creation and annihilation operators as
\beq
A_i =\sum_{j}\left(\cosh(g\lambda_{i})V_{ij}a_j+\sinh(g\lambda_{i})V_{ij} a_j^{\dagger}\right),\;\;\; \sqrt{\omega_i} = e^{g\lambda_{i}} .
\eeq
We can represent this Bogoliubov transformation in terms of conjugation by a unitary operator: 
\beq
A_i = \mathcal{U}^{\dagger} a_i\mathcal{U},\;\;\; \mathcal{U} = e^{\sum_i \frac{g\lambda_i}{2}\left(a_i^{\dagger}a_i^{\dagger} - a_ia_i\right)}e^{\sum_{i,j} v_{ij} a_i^{\dagger}a_j},
\eeq
where the real, anti-symmetric matrix $v_{ij}$ is defined as $V= e^{v}$. Therefore, the new vacuum $\psi$ in presence of the bilinear interaction can be related to the old vacuum $\psi_0$ as 
\beq
| \psi \rangle = \mathcal{U}^{\dagger} |\psi_0\rangle = e^{ \frac{1}{2}\sum_{j,k}B_{jk}\left(a_j^{\dagger}a_k^{\dagger} - a_ja_k\right)}|\psi_0\rangle,
\eeq
where
$$B_{jk} = g\sum_i \lambda_i V^T_{ji}V_{ik}.$$
We can also re-express this state in terms of the $\mathfrak{gl}(n,\mathbb{R})$ generators $\hat{\cO}_{ij} = \frac{1}{2}\left(\hat{\phi}_i\hat{\pi}_j + \hat{\pi}_j \hat{\phi}_i\right)$, which were discussed in Sec.~\ref{sec:HO}:
\beq
| \psi \rangle =  e^{2 i\sum_{i,j}B_{ij} \hat{\cO}_{ij} }|\psi_0\rangle,
\eeq
The binding complexity of the state can now be computed perturbatively in $g$, following our discussion in the previous section.  The leading order contribution is
\beq
\cC_b = 2 || \mathbf{B} || + \cdots =2\left( \sum_{i\neq j} B_{ij}B_{ji}\right)^{1/2} + \cdots,
\label{finalpertcomp}
\eeq
where $\mathbf{B} = \sum_{i, j} B_{ij} \hat{\cO}_{ij}$ and the $\cdots$ indicate higher order corrections which enter at $O(g^3)$ (as discussed in the previous section).

This result shows that adding ``double-trace deformations'' to the Hamiltonian creates binding complexity in the vacuum. If binding complexity measures the interior volume of wormholes, our result implies that the deformation has created a wormhole where none previously existed.   This is in analogy with the holographic results of \cite{Gao2017} where double-trace deformations of a product of CFTs enlarged a wormhole between the corresponding geometric asymptotic regions.   We computed our results above in a toy model of oscillators,  but we expect that a similar calculation will go through in the case of generalized free fields describing the large N limit of CFTs, which is the limit in which field theories are holographically described by classical geometry.

\section{Discussion}{\label{sec:discussion}}

\begin{figure}[htbp!]
\begin{center}
\includegraphics[width=.5\textwidth]{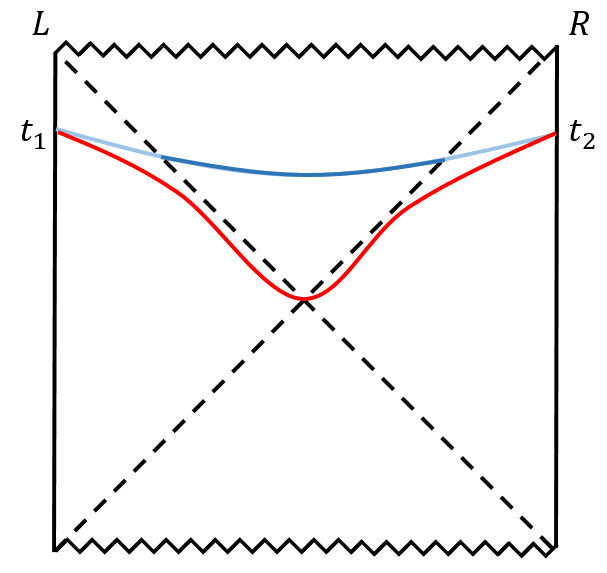}
\caption{Cauchy slices of maximal volume (blue) and of minimal interior volume (red) in the BTZ geometry, both anchored at boundary times $t_1$ in the left CFT and $t_2$ in the right CFT. The volume of the interior of the maximal slice (dark blue) increases over time, but the corresponding circuit does not minimize binding complexity. \label{fig:penrose}}
\end{center}
\end{figure}

We have suggested an interpretation for the volume of multiboundary wormhole interiors in AdS/CFT in terms of the binding complexity of the dual state. However, our discussion was limited to the interior volume of the time-reflection symmetric Cauchy slice in the bulk. If we consider a generic Cauchy surface ending at the times $(t_1, \cdots t_n)$ on the boundaries, then the volume of the wormhole interior will, in general, be larger. However, the binding complexity should be independent of the times $t_i$ because changing these times simply corresponds to local Hamiltonian evolution in the different CFTs, and does not add any entanglement. This observation suggests that the covariant version of the bulk dual to binding complexity should be given by minimizing the interior volume over all the bulk Cauchy surfaces \emph{and} over the different boundary times $\{t_i\}$. Note that if we consider the maximum volume slice in the bulk ending at the times $t_i$, then its volume is expected to be dual to the {\it total} complexity of the boundary state, which indeed depends on the $t_i$ because local Hamiltonian evolution adds to the total complexity. However, the corresponding circuit is \emph{not} the minimal one from the point of view of binding complexity. Fig.~\ref{fig:penrose} illustrates that the maximal volume Cauchy slice in the two-sided wormhole corresponding to the BTZ black hole can have a large interior volume, but it is always possible to find a different Cauchy slice that passes through the bifurcation surface.

The relation between binding complexity and wormhole interiors was most concrete for certain states created by performing the Euclidean path integral on a graph with locally bipartite connections between parties, but which can nevertheless have multipartite entanglement.  This occurs if some local degrees of freedom in each party have bipartite entanglement with local degrees of freedom in {\it different} parties.  This is a structure resembling the W-state on qubits \eqref{Wstate}.  However, we know that states with holographic duals satisfy the additional condition that mutual information is monogamous \cite{mmi}, implying that it is of the perfect tensor type \cite{chaos}.  In the bit-thread picture of entanglement, it seems necessary to sum over different bit-thread configurations to achieve this constraint \cite{headrick, cooperative}.  In our picture this would mean summing over multiple (perhaps all) Euclidean graphs that produce states on a given partition of external variables.   It would be interesting to consider the binding complexity for these kinds of states -- it is not obvious that the complexity will simply be a weighted sum of the complexities of the individual graph states.

Our notion of binding complexity has similarities to the idea of quantum communication complexity, where several independent parties attempt to collaborate on some particular computation.\footnote{We thank Scott Aaronson for bringing quantum communication complexity to our attention.} We can define the quantum communication complexity of a task to be the minimum number of qubits that must be exchanged between all the parties in order to complete the computation. Binding complexity measures a similar quantity, namely the number of gates that affect more than one party's qubits. In this way, both binding and quantum communication complexity increase as the computation requires more cooperation or interaction between the parties. In fact, we can obtain a strict relationship between the two quantities. Suppose all the gates in an $n$-qubit quantum circuit $U$ are $k$-qubit gates. Then the quantum communication complexity of applying $U$ to some distributed set of qubits is bounded by the binding complexity, since we may always transmit qubits across party lines in order to apply one of our gates. If the distributed parties run into a gate that contributes to the binding complexity during the application of $U$, they may simply communicate all the qubits to one of the involved parties, apply the unitary locally, and then send the qubits back to their proper owners (the bound is improved by a factor of 2 if we drop this last requirement). Each cross-boundary gate therefore contributes a maximum of $2k$ to the quantum communication complexity, and we obtain the upper bound
\begin{align}
C_{\text{qComm}}(U) \leq 2k C_b (U).
\end{align} 
Note that if we described this in a holographically dual geometry, the required multiboundary wormhole need not be traversable - there is no wormhole-based ``quantum FedEx" that would allow qubit transfers between the different boundaries, which we are treating as the distributed parties attempting to build the unitary $U$.
However, we can obtain a bound on the communication complexity of the problem by studying this geometry, assuming our conjecture holds. It would be interesting to make this analogy between binding and quantum communication complexity more precise in holography.

\subsection*{Acknowledgements}

We gratefully acknowledge useful discussions with Scott Aaronson, Matt Headrick, Matthew Hodel, Lampros Lamprou, Charles Rabideau, and G\'{a}bor S\'{a}rosi. MD is supported by the National Science Foundation Graduate Research Fellowship under Grant No. DGE-1845298. VB, OP, AK and MD were supported in part by the Simons Foundation (\# 385592, VB) through the It From Qubit Simons Collaboration, and the US Department of Energy contract \# FG02-05ER-41367. VB also acknowledges the hospitality of the Aspen Center for Physics which is supported by National Science Foundation grant PHY-1607611.

\appendix

\section{Binding complexity for more general states} \label{sec:ap2}

In this appendix, we compute the binding complexity for a state with less symmetry than that of (\ref{eq:omegacomp}). This will be an educational exercise that suggests a solution procedure for a totally arbitrary state. Consider a wavefunction for a four-party state taking the general Gaussian form (\ref{eq:graphgauss}) with the matrix $\Omega$ taking a block structure like
\begin{align}
\Omega = \begin{pmatrix} 
\omega \lambda_1 & \lambda_2^{(1)} & \lambda_2^{(2)} & \lambda_2^{(3)} \\ 
 \lambda_2^{(1)} & \omega \lambda_1 & \lambda_2^{(3)} & \lambda_2^{(2)} \\
 \lambda_2^{(2)} & \lambda_2^{(3)} & \omega \lambda_1 & \lambda_2^{(1)}  \\ 
 \lambda_2^{(3)} & \lambda_2^{(2)} &  \lambda_2^{(1)} & \omega \lambda_1 
\end{pmatrix}. \label{eq:omegadiff}
\end{align}
As in \eqref{eq:omegablock}, each entry above is an $N \times N$ matrix, where $N$ is the number of oscillators on each boundary. The elements $\lambda_2^{(i)}$ are the matrices all of whose elements are couplings similarly labeled $\lambda_2^{(i)}$ (below, $\lambda_2^{(1)}$ refers to the coupling, not the full matrix). The elements $\omega \lambda_1$ are matrices that are $\omega$ on the diagonal and $\lambda_1$ on all off-diagonals. This $\Omega$ is not completely general: in the language of Sec.~\ref{sec:pathint}, it corresponds to the path integral on a graph with $\mathbb{Z}_2 \times \mathbb{Z}_2$ symmetry between the four parties.

The solution of the Euler-Arnold equation (\ref{eq:ea}) is independent of the structure of the wavefunction, so the velocity matrix $V$ again is constant. In general, one can show that choosing the structure of the velocity matrix $V$ to have the same form as $\Omega$ will allow for solution of the flow equation (\ref{eq:matrixflow}). Consequently, we choose $V$ to take the same form as (\ref{eq:omegadiff}) with $a$ replacing $\omega$, $b$ replacing $\lambda_1$, and three cross-party velocities $c^1$, $c^2$, $c^3$ replacing $\lambda_2^{(1)}, \lambda_2^{(2)},\lambda_2^{(3)}$. Doing so, (\ref{eq:matrixflow}) splits into a $5 \times 5$ matrix equation:
\begin{align}
\frac{d\vec{\Omega}}{ds}  = M\vec{\Omega}, \label{eq:matrixeqn}
\end{align}
where $\vec{\Omega} = \begin{pmatrix} \omega &\lambda_1 & \lambda_2^{(1)} & \lambda_2^{(2)} & \lambda_2^{(3)} \end{pmatrix}^T$ arranges the $s$-dependent couplings of the matrix $\Omega$ into a vector and
\begin{align}
M = 2
\begin{pmatrix}
a & (N-1) b & Nc^1 & Nc^2 & Nc^3 \\
b & a + (N-2) b &Nc^1 & Nc^2 & Nc^3 \\
c^1 &(N-1)c^1 &a+(N-1)b & Nc^3& Nc^2 \\
c^2 &(N-1)c^2&Nc^3&a+(N-1)b&Nc^1\\
c^3&(N-1)c^3 &Nc^2&Nc^1 & a+(N-1)b
\end{pmatrix}.
\end{align}
For comparison, note that the equations (\ref{eq:couplesystem1}) - (\ref{eq:couplesystem3}) can be written as a similar $3 \times 3$ matrix equation. The matrix $M$ has five distinct eigenvalues:
\begin{align}
\kappa_0 &=2( a-b) \\
\kappa_1 &= 2(a+b(N-1) + (c_1 - c_2 - c_3)N) \\
\kappa_2 &= 2(a+b(N-1)+(-c_1 + c_2 - c_3)N)\\
\kappa_3 &=2( a+b(N-1) + (-c_1 - c_2 + c_3)N)\\
\kappa_+ &=2( a+b(N-1) + (c_1 + c_2 + c_3)N).
\end{align}
Solving (\ref{eq:matrixeqn}) with the usual boundary conditions of $\Omega^{(i)} =  \text{diag} (\omega_0, \omega_0, \ldots, \omega_0)$ at $s=0$ and $\Omega^{(f)}$ as given by (\ref{eq:omegadiff}) at $s=1$, we find
\begin{align}
\omega &=  \frac{\omega_0}{4N} (4(N-1)e^{\kappa_0 } + e^{\kappa_1} + e^{\kappa_2} + e^{\kappa_3} + e^{\kappa_+}) \label{eq:eigen1} \\
\lambda_1 &= -\frac{\omega_0}{4N} (4e^{\kappa_0} - e^{\kappa_1} - e^{\kappa_2} - e^{\kappa_3} - e^{\kappa_+})\\
\lambda_2^{(1)} &= \frac{\omega_0}{4N} (e^{\kappa_1} - e^{\kappa_2} - e^{\kappa_3} + e^{\kappa_+}) \\
\lambda_2^{(2)} &= -\frac{\omega_0}{4N} (e^{\kappa_1} - e^{\kappa_2} + e^{\kappa_3} - e^{\kappa_+}) \\
\lambda_2^{(3)} &= -\frac{\omega_0}{4N} (e^{\kappa_1} + e^{\kappa_2} - e^{\kappa_3} - e^{\kappa_+}).\label{eq:eigen5} 
\end{align}
We remark that the five distinct eigenvalues of $\Omega$ are given by:
\begin{align}
\rho_0 &= \omega - \lambda_1 \\
\rho_1 &= \omega + (N-1)\lambda_1 + N(\lambda_2^{(1)} - \lambda_2^{(2)} - \lambda_2^{(3)}) \\
\rho_2 &= \omega + (N-1)\lambda_1 + N(-\lambda_2^{(1)} + \lambda_2^{(2)} - \lambda_2^{(3)}) \\
\rho_3 &= \omega + (N-1)\lambda_1 + N(-\lambda_2^{(1)} - \lambda_2^{(2)} + \lambda_2^{(3)})\\
\rho_+ &= \omega + (N-1)\lambda_1 + N(\lambda_2^{(1)} + \lambda_2^{(2)} + \lambda_2^{(3)}),
\end{align}
closely related to the eigenvalues of $M$. In the permutation-symmetric limit, $\rho_+$ corresponds to $\lambda_+$, $\rho_0$ corresponds to $\lambda_0$, and $\rho_1, \rho_2, \rho_3$ all approach $\lambda_-$. Solving the system (\ref{eq:eigen1}) - (\ref{eq:eigen5}) for the eigenvalues $\kappa$ we obtain
\begin{align}
\kappa_i &= \ln(\frac{\rho_i}{\omega_0}),
\end{align}
for all $i =0,1,2,3,+$. Finally, solving for the velocities $a,b,c^k$,
\begin{align}
a &= 4(N-1) \kappa_0 + \kappa_1 + \kappa_2 + \kappa_3+ \kappa_+ \\
b&= -4\kappa_0 + \kappa_1 + \kappa_2 + \kappa_3+ \kappa_+ \\
c^1 &= \kappa_1 - \kappa_2 - \kappa_3 + \kappa_+ \\
c^2 &= -\kappa_1 + \kappa_2 - \kappa_3 + \kappa_+ \\
c^3 &= -\kappa_1 - \kappa_2 + \kappa_3 + \kappa_+.
\end{align}
Rewriting the $c^k$ that determine the binding complexity in terms of the eigenvalues $\rho_i$,
\begin{align}
c^1 &= \frac{1}{8N} \ln \frac{\rho_1 \rho_+}{\rho_2 \rho_3} \\
c^2 &=\frac{1}{8N} \ln \frac{\rho_2 \rho_+}{\rho_1 \rho_3} \\
c^3 &= \frac{1}{8N}\ln \frac{\rho_3 \rho_+}{\rho_1 \rho_2} .
\end{align}
Lastly, a short combinatorial computation determines the binding complexity
\begin{align}
\mathcal{C}_b = 2N|c|,
\end{align}
where $|c| = \sqrt{(c^1)^2+(c^2)^2+(c^3)^2}$. Notice that again the prefactor of $N$ above cancels the $N$ dependence of the $c^k$ so that the binding complexity is finite in the large $N$ limit. 

Unfortunately, the binding complexity does not arrange nicely in terms of a parameter $\mu$ as in Sec.~\ref{sec:mbycomplex}, and it is prohibitively diffcult to evaluate the entanglement entropy associated with a single party of the state specified by (\ref{eq:omegadiff}) to obtain a complexity-entropy scaling. Nevertheless, this computation is instructive to understand how to compute the binding complexity for a (more) general Gaussian state. In general, we expect that if we arrange the couplings in $\Omega$ into a vector $\vec{\Omega}$, the eigenvalues $\vec{\rho}$ of the matrix $\Omega$ will be some linear combination of the couplings: $\vec{\rho} = A\vec{\Omega}$. In this case, choosing $V$ to have the same matrix structure as $\Omega$ gives rise to a lower-dimensional matrix equation for the couplings in terms of a matrix $M = 2A$. The eigenvalues $\vec{\kappa}$ of $M$ will be $\vec{\kappa} = M\vec{V}$, and the solution for the velocities will looks like $V_i = \frac12 \sum_j (A^{-1})_{ij} \ln \frac{\rho_j}{\omega_0}$. Note that $\vec{V}$ is the vector of velocities analogous to $\vec{\Omega}$. Once the velocities are obtained, it is straightforward to compute the binding complexity based on the particular combinatorics of a given setup.

\section{Wavefunctions of permutation-symmetric graphs} \label{sec:ap1}

In this appendix, we compute the wavefunctions of the branched graphs presented in Fig.~\ref{fig:branch1}, for an arbitrary number of parties $m$ and number of oscillators per party $N$, working in the $M \to 0 $ limit for simplicity. In this limit, the propagator \eqref{MK} remains Gaussian and takes the simple form
\begin{align}
K (x_1, x_2, \beta) \propto  e^{-\frac{1}{2\beta} (x_2 - x_1)^2}.
\end{align}
Starting from the definition (\ref{eq:pathintproc}) and reading off from Fig.~\ref{fig:branch1} where each oscillator vertex connects to internal vertices,
\begin{align}
\psi(\vec{x}) &= \int d\vec{y} \prod_{(v_1,v_2,T) \in E_G} K(v_1,v_2,T) \\
&=  \tilde{N} \int d\vec{y} dy_c \exp \biggl[-\frac{1}{2T_1} \left((x_1^1 - y_1)^2 + \ldots   + (x_m^N-y_m)^2 \right) \nonumber\\
&\qquad\qquad\qquad\qquad - \frac{1}{2T_2} \left((y_1 - y_c)^2 + \ldots + (y_m - y_c)^2 \right) \biggr] \\
&= \tilde{N} \int d\vec{y}\exp \left[-\frac{1}{2T_1} \left(N\sum_i y_i^2 + \sum_{i,j} (x_j^i)^2 - 2\sum_j y_j \sum_i x_j^i \right) \right] \nonumber\\
&\qquad\times \int dy_c \exp \left[- \frac{1}{2T_2} \left(my_c^2 -2y_c \sum_i y_i +\sum_i y_i^2 \right) \right] ,
\end{align}
where $\tilde{N}$ is a normalization constant. Performing the Gaussian integral over $y_c$, we obtain
\begin{align}
\psi(\vec{x}) &= \tilde{N}' \int d\vec{y}\exp\biggl[-\frac{1}{2T_1} \left(N\sum_i y_i^2 + \sum_{i,j} (x_j^i)^2 - 2\sum_j y_j \sum_i x_j^i \right) \nonumber \\
&\qquad\qquad\qquad\qquad - \frac{1}{2T_2} \left(\sum_i y_i^2  - \frac{1}{m} (\sum_i y_i)^2\right) \biggr], \label{eq:remain}
\end{align}
where $\tilde{N}'$ is a new normalization constant. The remaining integral (\ref{eq:remain}) is also Gaussian over the internal vertices $y_i$, although it has a linear term. That is, it takes the form
\begin{align}
\psi(\vec{x}) &= \tilde{N}'  \exp\left[-\frac{1}{2T_1} \sum_{i,j}(x_j^i)^2 \right] \int d\vec{y} \exp\left[-\frac12 \vec{y}^T A \vec{y} + \vec{B}^T \vec{y}\right], \label{eq:gaussianint}
\end{align}
where the matrix $A$ and vector $\vec{B}^T$ are given by:
\begin{align}
A_{ij} = \alpha \delta_{ij} + \beta (1-\delta_{ij}), \qquad B^T_j = \frac{1}{T_1} \sum_i x_j^i.
\end{align}
That is, $A$ takes value $\alpha $ on the diagonal and $\beta$ on all off-diagonals. The constants $\alpha$ and $\beta$ are given in terms of $T_1$, $T_2$, $m$, and $N$ by
\begin{align}
\alpha = \frac{N}{T_1} + \frac{1}{T_2} \left(1-\frac{1}{m}\right), \qquad \beta= -\frac{1}{mT_2}.
\end{align}
The exact solution of the general Gaussian matrix integral of the form of (\ref{eq:gaussianint}) is well-known. Evaluating it gives
\begin{align}
\psi(\vec{x}) &= \tilde{N}''  \exp\left[-\frac{1}{2T_1} \sum_{i,j}(x_j^i)^2 \right]\exp\left[\frac12 \vec{B}^T A^{-1} \vec{B} \right] , 
\end{align}
where $\tilde{N}''$ is another new normalization constant. The inverse of $A$ has the same symmetry as $A$, with $A^{-1}_{ij} = P \delta_{ij} + Q(1-\delta_{ij})$ and
\begin{align}
P = \frac{\alpha+(m-2)\beta}{ \alpha^2 + (m-2)\alpha \beta-(m-1)\beta^2}, \qquad Q = \frac{-\beta}{ \alpha^2 + (m-2)\alpha \beta-(m-1)\beta^2}.
\end{align}
Therefore, in terms of the oscillator variables $x_j^i$, the wavefunction is
\begin{align}
\psi (\vec{x}) = \tilde{N}'' \exp\left[-\frac{1}{2T_1} \sum_{i,j}(x_j^i)^2 \right]\exp\left[\frac{1}{2T_1^2}  \left(P \sum_j (\sum_i x_j^i)^2 + Q \sum_{j\neq k} \sum_{i, \ell} x_j^i x_k^{\ell} \right)\right]. \label{eq:wavefun}
\end{align}
Despite the cumbersome sum notation for the general case, one can check that this is indeed Gaussian and can be written in the standard Gaussian form $\psi (\vec{x})  = \tilde{N}'' \exp(-\frac12 \vec{x}^T \Omega \vec{x})$ with $\Omega$ in the form of (\ref{eq:omegacomp}). To be completely explicit, $\Omega$ has the general permutation-symmetric form (\ref{eq:omegacomp}) with
\begin{align}
\omega &= \frac{1}{T_1} - \frac{P}{T_1^2} = \frac{T_1 (mN - 1) + T_2 mN(N-1)}{T_1 mN (T_1 + NT_2)} \\
\lambda_1 &= -\frac{P}{T_1^2} =-\frac{T_1 +mNT_2}{T_1 mN (T_1 + NT_2)}  \\
\lambda_2 &= -\frac{Q}{T_1^2} =-\frac{T_1 }{T_1 mN (T_1 + NT_2)} .
\end{align}
in agreement with the $M \to 0$ limit of (\ref{eq:omegcouple}) - (\ref{eq:massivecouple}). This computation was entirely in the $M \to 0$ limit, but the trick employed herein of rewriting the product over propagators as matrix Gaussian integrals works very generally. For any permutation-symmetric graph the computation goes through identically with possibly different values of $\alpha$ and $\beta$, even when $M \neq 0$.

\section{Perturbation theory to $\mathcal{O}(t^3)$}\label{app3}
For completeness, we will show how to proceed at $O(t^3)$ in the small-time perturbation theory in this appendix. At third order, we find
\beq
v_{(3)}^J(s) =  v_{(3)}^J(0)+\frac{s(s-1)}{2} \sum_{K,L,M,N} \left({c_{KL}}^J{c_{MN}}^L+{c_{LK}}^J{c_{MN}}^L\right)h^Kh^Mh^N.
\eeq
So, now the unitary becomes
\beqn
U(1) &=& \mathcal{P}\exp\left(i \int_0^1 ds \,\sum_I \left[t h^I - \frac{1}{2}t^2\left(1 -2s\right)\sum_{K,L} {c_{KL}}^Ih^Kh^L + t^3v_{(3)}^I(0)\right.\right.\\
&\quad& +\left.\left.t^3\frac{s(s-1)}{2} \sum_{K,L,M,N} \left({c_{KL}}^I{c_{MN}}^L+{c_{LK}}^I{c_{MN}}^L\right)h^Kh^Mh^N\right]\cO_I+\cdots\right)\nonumber\\
&=& 1 +i\sum_I\left[t h^I +t^3 v_{(3)}^I(0)- \frac{t^3}{24} \sum_{K,L,M,N} \left({c_{KL}}^I{c_{MN}}^L+{c_{LK}}^I{c_{MN}}^L\right)h^Kh^Mh^N \right]\cO_I \nonumber\\
&\quad& -\frac{t^2}{2} \sum_{I,J} h^Ih^J\cO_I\cO_J +\frac{1}{2}t^3 \sum_{I,J,K,L} \int_0^1ds_1\int_0^1ds_2 \biggl[\Theta(s_1-s_2) \cO_I\cO_J(\frac{1}{2}-s_2) \nonumber \\
&\quad& +\Theta(s_2-s_1) \cO_J\cO_I (\frac{1}{2}-s_2)\biggr]{c_{KL}}^Jh^Ih^Kh^L- \frac{it^3}{3!} \sum_{I,J,K} h^Ih^Jh^K \cO_I\cO_J\cO_K + \cdots \nonumber\\
&=& 1 +i\sum_I\left[t h^I  +t^3 v_{(3)}^I(0)- \frac{t^3}{24} \sum_{K,L,M,N} \left({c_{KL}}^I{c_{MN}}^L+{c_{LK}}^I{c_{MN}}^L\right) h^Kh^Mh^N \right]\cO_I \nonumber \\
&\quad& -\frac{t^2}{2} \sum_{I,J} h^Ih^J\cO_I\cO_J +\frac{t^3}{24} \sum_{I,J,K,L}  \left[ \cO_I\cO_J- \cO_J\cO_I \right]{c_{KL}}^Jh^Ih^Kh^L \nonumber \\
&\quad& -\frac{it^3}{3!} \sum_{I,J,K} h^Ih^Jh^K \cO_I\cO_J\cO_K + \cdots\nonumber\\
&=& 1 +i \sum_I \biggl[t h^I  +t^3 v_{(3)}^I(0)- \frac{t^3}{24}  \sum_{K,L,M,N} \bigl({c_{KL}}^I{c_{MN}}^L+{c_{LK}}^I{c_{MN}}^L \nonumber \\
&\quad& +{f_{KL}}^I{c_{MN}}^L\bigr) h^Kh^Mh^N \biggr]\cO_I  -\frac{t^2}{2} \sum_{I,J} h^Ih^J\cO_I\cO_J-\frac{it^3}{3!} \sum_{I,J,K} h^Ih^Jh^K \cO_I\cO_J\cO_K + \cdots.\nonumber
\eeqn
Comparing with equation \eqref{bc}, this implies
\beq
v_{(3)}^I(0)= \frac{1}{24} \sum_{K,L,M,N} \left({c_{KL}}^I{c_{MN}}^L+{c_{LK}}^I{c_{MN}}^L+{f_{KL}}^I{c_{MN}}^L\right)h^Kh^Mh^N.
\eeq
So the total velocity at this order is given by
\beqn
v^I(s) &=& t h^I+ \frac{1}{2}t^2\left(1-2s\right) \sum_{K,L} {c_{KL}}^Ih^Kh^L+  \frac{t^3}{24} \sum_{K,L,M,N} \Big\{ \bigl({c_{KL}}^I{c_{MN}}^L+{c_{LK}}^I{c_{MN}}^L \nonumber \\
&\quad& +{f_{KL}}^I{c_{MN}}^L\bigr) +\frac{s(s-1)}{2} \left({c_{KL}}^I{c_{MN}}^L+{c_{LK}}^I{c_{MN}}^L\right)\Big\}h^Kh^Mh^N + O(t^4). 
\eeqn
We can now compute the complexity at $O(t^3)$, if we so desire. 

\providecommand{\href}[2]{#2}\begingroup\raggedright\endgroup


\begin{thebibliography}{10}

\bibitem{quant-ph/9508027}
P.~W. Shor, \emph{Polynomial-time algorithms for prime factorization and
  discrete logarithms on a quantum computer},
  \href{http://dx.doi.org/10.1137/S0097539795293172}{\emph{SIAM Journal on
  Computing} {\bf 26} (1997) 1484--1509},
  [\href{https://arxiv.org/abs/arXiv:quant-ph/9508027}{{\tt
  arXiv:quant-ph/9508027}}].

\bibitem{PhysRevA.62.062314}
W.~D\"ur, G.~Vidal and J.~I. Cirac, \emph{Three qubits can be entangled in two
  inequivalent ways},
  \href{http://dx.doi.org/10.1103/PhysRevA.62.062314}{\emph{Phys. Rev. A} {\bf
  62} (Nov, 2000) 062314},
  [\href{https://arxiv.org/abs/arXiv:quant-ph/0005115}{{\tt
  arXiv:quant-ph/0005115}}].

\bibitem{PhysRevA.61.052306}
V.~Coffman, J.~Kundu and W.~K. Wootters, \emph{Distributed entanglement},
  \href{http://dx.doi.org/10.1103/PhysRevA.61.052306}{\emph{Phys. Rev. A} {\bf
  61} (Apr, 2000) 052306},
  [\href{https://arxiv.org/abs/arXiv:quant-ph/9907047}{{\tt
  arXiv:quant-ph/9907047}}].

\bibitem{1606.09290}
F.~Diker, \emph{Deterministic construction of arbitrary w states with
  quadratically increasing number of two-qubit gates},
  \href{https://arxiv.org/abs/arXiv:1606.09290}{{\tt arXiv:1606.09290}}.

\bibitem{Jefferson2017}
R.~Jefferson and R.~C. Myers, \emph{Circuit complexity in quantum field
  theory}, \href{http://dx.doi.org/10.1007/JHEP10(2017)107}{\emph{Journal of
  High Energy Physics} {\bf 2017} (Oct, 2017) 107},
  [\href{https://arxiv.org/abs/arXiv:1707.08570}{{\tt arXiv:1707.08570}}].

\bibitem{quant-ph/0502070}
M.~A. Nielsen, \emph{A geometric approach to quantum circuit lower bounds},
  \href{https://arxiv.org/abs/arXiv:quant-ph/0502070}{{\tt
  arXiv:quant-ph/0502070}}.

\bibitem{quant-ph/0603161}
M.~A. Nielsen, M.~R. Dowling, M.~Gu and A.~C. Doherty, \emph{Quantum
  computation as geometry},
  \href{http://dx.doi.org/10.1126/science.1121541}{\emph{Science} {\bf 311}
  (2006) 1133--1135}, [\href{https://arxiv.org/abs/arXiv:quant-ph/0603161}{{\tt
  arXiv:quant-ph/0603161}}].

\bibitem{quant-ph/0701004}
M.~R. Dowling and M.~A. Nielsen, \emph{The geometry of quantum computation},
  \href{https://arxiv.org/abs/arXiv:quant-ph/0701004}{{\tt
  arXiv:quant-ph/0701004}}.

\bibitem{fermion1}
R.~Khan, C.~Krishnan and S.~Sharma, \emph{Circuit complexity in fermionic field
  theory},  \href{https://arxiv.org/abs/arXiv:1801.07620}{{\tt
  arXiv:1801.07620}}.

\bibitem{fermion2}
L.~Hackl and R.~C. Myers, \emph{Circuit complexity for free fermions},
  \href{http://dx.doi.org/10.1007/JHEP07(2018)139}{\emph{Journal of High Energy
  Physics} {\bf 2018} (Jul, 2018) 139},
  [\href{https://arxiv.org/abs/arXiv:1803.10638}{{\tt arXiv:1803.10638}}].

\bibitem{coherent}
M.~Guo, J.~Hernandez, R.~C. Myers and S.-M. Ruan, \emph{Circuit complexity for
  coherent states},  \href{https://arxiv.org/abs/arXiv:1807.07677}{{\tt
  arXiv:1807.07677}}.

\bibitem{interacting}
A.~Bhattacharyya, A.~Shekar and A.~Sinha, \emph{Circuit complexity in
  interacting qfts and rg flows},
  \href{https://arxiv.org/abs/arXiv:1808.03105}{{\tt arXiv:1808.03105}}.

\bibitem{quench1}
H.~A. Camargo, P.~Caputa, D.~Das, M.~P. Heller and R.~Jefferson,
  \emph{Complexity as a novel probe of quantum quenches: universal scalings and
  purifications},  \href{https://arxiv.org/abs/arXiv:1807.07075}{{\tt
  arXiv:1807.07075}}.

\bibitem{quench2}
D.~W.~F. Alves and G.~Camilo, \emph{Evolution of complexity following a quantum
  quench in free field theory},
  \href{http://dx.doi.org/10.1007/JHEP06(2018)029}{\emph{Journal of High Energy
  Physics} {\bf 2018} (Jun, 2018) 29},
  [\href{https://arxiv.org/abs/arXiv:1804.00107}{{\tt arXiv:1804.00107}}].

\bibitem{hamilcomplex}
R.-Q. Yang and K.-Y. Kim, \emph{Complexity of operators generated by quantum
  mechanical hamiltonians},  \href{https://arxiv.org/abs/arXiv:1810.09405}{{\tt
  arXiv:1810.09405}}.

\bibitem{PhysRevD.97.066004}
R.-Q. Yang, \emph{Complexity for quantum field theory states and applications
  to thermofield double states},
  \href{http://dx.doi.org/10.1103/PhysRevD.97.066004}{\emph{Phys. Rev. D} {\bf
  97} (Mar, 2018) 066004}, [\href{https://arxiv.org/abs/arXiv:1709.00921}{{\tt
  arXiv:1709.00921}}].

\bibitem{1710.00600}
K.-Y. Kim, C.~Niu, R.-Q. Yang and C.-Y. Zhang, \emph{Comparison of holographic
  and field theoretic complexities by time dependent thermofield double
  states},  \href{https://arxiv.org/abs/arXiv:1710.00600}{{\tt
  arXiv:1710.00600}}.

\bibitem{1803.01797}
R.-Q. Yang, Y.-S. An, C.~Niu, C.-Y. Zhang and K.-Y. Kim, \emph{Principles and
  symmetries of complexity in quantum field theory},
  \href{https://arxiv.org/abs/arXiv:1803.01797}{{\tt arXiv:1803.01797}}.

\bibitem{1809.06678}
R.-Q. Yang, Y.-S. An, C.~Niu, C.-Y. Zhang and K.-Y. Kim, \emph{More on
  complexity of operators in quantum field theory},
  \href{https://arxiv.org/abs/arXiv:1809.06678}{{\tt arXiv:1809.06678}}.

\bibitem{complexityqft2}
S.~Chapman, M.~P. Heller, H.~Marrochio and F.~Pastawski, \emph{Towards
  complexity for qantum field theory states},
  \href{https://arxiv.org/abs/arXiv:1707.08582}{{\tt arXiv:1707.08582}}.

\bibitem{bartek}
B.~Czech, \emph{Einstein equations from varying complexity},
  \href{http://dx.doi.org/10.1103/PhysRevLett.120.031601}{\emph{Phys. Rev.
  Lett.} {\bf 120} (Jan, 2018) 031601},
  [\href{https://arxiv.org/abs/arXiv:1706.00965}{{\tt arXiv:1706.00965}}].

\bibitem{kyoto1}
P.~Caputa, N.~Kundu, M.~Miyaji, T.~Takayanagi and K.~Watanabe, \emph{Anti{-}de
  sitter space from optimization of path integrals in conformal field
  theories},
  \href{http://dx.doi.org/10.1103/PhysRevLett.119.071602}{\emph{Phys. Rev.
  Lett.} {\bf 119} (Aug, 2017) 071602},
  [\href{https://arxiv.org/abs/arXiv:1703.00456}{{\tt arXiv:1703.00456}}].

\bibitem{kyoto2}
P.~Caputa, N.~Kundu, M.~Miyaji, T.~Takayanagi and K.~Watanabe, \emph{Liouville
  action as path-integral complexity: from continuous tensor networks to
  ads/cft}, \href{http://dx.doi.org/10.1007/JHEP11(2017)097}{\emph{Journal of
  High Energy Physics} {\bf 2017} (Nov, 2017) 97},
  [\href{https://arxiv.org/abs/arXiv:1706.07056}{{\tt arXiv:1706.07056}}].

\bibitem{Bhattacharyya2018}
A.~Bhattacharyya, P.~Caputa, S.~R. Das, N.~Kundu, M.~Miyaji and T.~Takayanagi,
  \emph{Path-integral complexity for perturbed cfts},
  \href{http://dx.doi.org/10.1007/JHEP07(2018)086}{\emph{Journal of High Energy
  Physics} {\bf 2018} (Jul, 2018) 86},
  [\href{https://arxiv.org/abs/arXiv:1804.01999}{{\tt arXiv:1804.01999}}].

\bibitem{1808.09072}
T.~Takayanagi, \emph{Holographic spacetimes as quantum circuits of
  path-integrations},  \href{https://arxiv.org/abs/arXiv:1808.09072}{{\tt
  arXiv:1808.09072}}.

\bibitem{Molina-Vilaplana2018}
J.~Molina-Vilaplana and A.~del Campo, \emph{Complexity functionals and
  complexity growth limits in continuous mera circuits},
  \href{http://dx.doi.org/10.1007/JHEP08(2018)012}{\emph{Journal of High Energy
  Physics} {\bf 2018} (Aug, 2018) 12},
  [\href{https://arxiv.org/abs/arXiv:1803.02356}{{\tt arXiv:1803.02356}}].

\bibitem{Magan2018}
J.~M. Mag{\'a}n, \emph{Black holes, complexity and quantum chaos},
  \href{http://dx.doi.org/10.1007/JHEP09(2018)043}{\emph{Journal of High Energy
  Physics} {\bf 2018} (Sep, 2018) 43},
  [\href{https://arxiv.org/abs/arXiv:1805.05839}{{\tt arXiv:1805.05839}}].

\bibitem{1807.04422}
P.~Caputa and J.~M. Magan, \emph{Quantum computation as gravity},
  \href{https://arxiv.org/abs/arXiv:1807.04422}{{\tt arXiv:1807.04422}}.

\bibitem{complexitytime}
T.~Ali, A.~Bhattacharyya, S.~S. Haque, E.~H. Kim and N.~Moynihan, \emph{Time
  evolution of complexity: A critique of three methods},
  \href{https://arxiv.org/abs/arXiv:1810.02734}{{\tt arXiv:1810.02734}}.

\bibitem{arnold}
V.~Arnold, \emph{Sur la géométrie différentielle des groupes de lie de
  dimension infinie et ses applications à l'hydrodynamique des fluides
  parfaits}, \href{http://dx.doi.org/10.5802/aif.233}{\emph{Annales de
  l'Institut Fourier} {\bf 16} (1966) 319--361}.

\bibitem{terrytao}
T.~Tao, \emph{The euler-arnold equation},
  {\emph{https://terrytao.wordpress.com/2010/06/07/the-euler-arnold-equation/}
  (2010) }.

\bibitem{ERepr}
J.~Maldacena and L.~Susskind, \emph{Cool horizons for entangled black holes},
  \href{http://dx.doi.org/10.1002/prop.201300020}{\emph{Fortschritte der
  Physik} {\bf 61} (2013) 781--811},
  [\href{https://arxiv.org/abs/arXiv:1306.0533}{{\tt arXiv:1306.0533}}].

\bibitem{Vijay2014}
V.~Balasubramanian, P.~Hayden, A.~Maloney, D.~Marolf and S.~F. Ross,
  \emph{Multiboundary wormholes and holographic entanglement},
  \href{http://dx.doi.org/10.1088/0264-9381/31/18/185015}{\emph{Classical and
  Quantum Gravity} {\bf 31} (2014) 185015},
  [\href{https://arxiv.org/abs/arXiv:1406.2663}{{\tt arXiv:1406.2663}}].

\bibitem{1406.2678}
D.~Stanford and L.~Susskind, \emph{Complexity and shock wave geometries},
  \href{http://dx.doi.org/10.1103/PhysRevD.90.126007}{\emph{Phys. Rev. D} {\bf
  90} (Dec, 2014) 126007}, [\href{https://arxiv.org/abs/arXiv:1406.2678}{{\tt
  arXiv:1406.2678}}].

\bibitem{1402.5674}
L.~Susskind, \emph{Computational complexity and black hole horizons},
  \href{https://arxiv.org/abs/arXiv:1402.5674}{{\tt arXiv:1402.5674}}.

\bibitem{1403.5695}
L.~Susskind, \emph{Addendum to computational complexity and black hole
  horizons},  \href{https://arxiv.org/abs/arXiv:1403.5695}{{\tt
  arXiv:1403.5695}}.

\bibitem{1408.2823}
L.~Susskind and Y.~Zhao, \emph{Switchbacks and the bridge to nowhere},
  \href{https://arxiv.org/abs/arXiv:1408.2823}{{\tt arXiv:1408.2823}}.

\bibitem{Brill1}
D.~R. Brill, \emph{Multi-black-hole geometries in (2+1)-dimensional gravity},
  \href{http://dx.doi.org/10.1103/PhysRevD.53.R4133}{\emph{Phys. Rev. D} {\bf
  53} (Apr, 1996) R4133--R4137},
  [\href{https://arxiv.org/abs/arXiv:gr-qc/9511022}{{\tt
  arXiv:gr-qc/9511022}}].

\bibitem{Brill2}
D.~R. Brill, \emph{Black holes and wormholes in 2+1 dimensions},
  \href{https://arxiv.org/abs/arXiv:gr-qc/9904083}{{\tt arXiv:gr-qc/9904083}}.

\bibitem{Brill3}
S.~Åminneborg, I.~Bengtsson, D.~Brill, S.~Holst and P.~Peldán, \emph{Black
  holes and wormholes in 2+1 dimensions}, {\emph{Classical and Quantum Gravity}
  {\bf 15} (1998) 627}, [\href{https://arxiv.org/abs/arXiv:gr-qc/9707036}{{\tt
  arXiv:gr-qc/9707036}}].

\bibitem{Skenderis2011}
K.~Skenderis and B.~C. van Rees, \emph{Holography and wormholes in 2+1
  dimensions},
  \href{http://dx.doi.org/10.1007/s00220-010-1163-z}{\emph{Communications in
  Mathematical Physics} {\bf 301} (Feb, 2011) 583--626},
  [\href{https://arxiv.org/abs/arXiv:0912.2090}{{\tt arXiv:0912.2090}}].

\bibitem{Krasnov1}
K.~Krasnov, \emph{Holography and riemann surfaces}, {\emph{Advances in
  Theoretical and Mathematical Physics} {\bf 4} (2000) 929--979},
  [\href{https://arxiv.org/abs/arXiv:hep-th/0005106}{{\tt
  arXiv:hep-th/0005106}}].

\bibitem{Krasnov2}
K.~Krasnov, \emph{Black-hole thermodynamics and riemann surfaces},
  {\emph{Classical and Quantum Gravity} {\bf 20} (2003) 2235},
  [\href{https://arxiv.org/abs/arXiv:gr-qc/0302073}{{\tt
  arXiv:gr-qc/0302073}}].

\bibitem{marolf}
D.~Marolf, H.~Maxfield, A.~Peach and S.~Ross, \emph{Hot multiboundary wormholes
  from bipartite entanglement}, {\emph{Classical and Quantum Gravity} {\bf 32}
  (2015) 215006}, [\href{https://arxiv.org/abs/arXiv:1506.04128}{{\tt
  arXiv:1506.04128}}].

\bibitem{Fu2018}
Z.~Fu, A.~Maloney, D.~Marolf, H.~Maxfield and Z.~Wang, \emph{Holographic
  complexity is nonlocal},
  \href{http://dx.doi.org/10.1007/JHEP02(2018)072}{\emph{Journal of High Energy
  Physics} {\bf 2018} (Feb, 2018) 72},
  [\href{https://arxiv.org/abs/arXiv:1801.01137}{{\tt arXiv:1801.01137}}].

\bibitem{ross}
A.~Peach and S.~F. Ross, \emph{Tensor network models of multiboundary
  wormholes}, {\emph{Classical and Quantum Gravity} {\bf 34} (2017) 105011},
  [\href{https://arxiv.org/abs/arXiv:1702.05984}{{\tt arXiv:1702.05984}}].

\bibitem{Gao2017}
P.~Gao, D.~L. Jafferis and A.~C. Wall, \emph{Traversable wormholes via a double
  trace deformation},
  \href{http://dx.doi.org/10.1007/JHEP12(2017)151}{\emph{Journal of High Energy
  Physics} {\bf 2017} (Dec, 2017) 151},
  [\href{https://arxiv.org/abs/arXiv:1608.05687}{{\tt arXiv:1608.05687}}].

\bibitem{2003PhRvA..67e2301N}
M.~A. {Nielsen}, C.~M. {Dawson}, J.~L. {Dodd}, A.~{Gilchrist}, D.~{Mortimer},
  T.~J. {Osborne} et~al., \emph{{Quantum dynamics as a physical resource}},
  \href{http://dx.doi.org/10.1103/PhysRevA.67.052301}{\emph{Phys. Rev. A} {\bf
  67} (May, 2003) 052301}, [\href{https://arxiv.org/abs/quant-ph/0208077}{{\tt
  quant-ph/0208077}}].

\bibitem{1996PhRvL..77.1413P}
A.~{Peres}, \emph{{Separability Criterion for Density Matrices}},
  \href{http://dx.doi.org/10.1103/PhysRevLett.77.1413}{\emph{Physical Review
  Letters} {\bf 77} (Aug., 1996) 1413--1415},
  [\href{https://arxiv.org/abs/quant-ph/9604005}{{\tt quant-ph/9604005}}].

\bibitem{1996PhLA..223....1H}
M.~{Horodecki}, P.~{Horodecki} and R.~{Horodecki}, \emph{{Separability of mixed
  states: necessary and sufficient conditions}},
  \href{http://dx.doi.org/10.1016/S0375-9601(96)00706-2}{\emph{Physics Letters
  A} {\bf 223} (Feb., 1996) 1--8},
  [\href{https://arxiv.org/abs/quant-ph/9605038}{{\tt quant-ph/9605038}}].

\bibitem{2002PhRvA..65c2314V}
G.~{Vidal} and R.~F. {Werner}, \emph{{Computable measure of entanglement}},
  \href{http://dx.doi.org/10.1103/PhysRevA.65.032314}{\emph{Phys. Rev. A} {\bf
  65} (Mar., 2002) 032314}, [\href{https://arxiv.org/abs/quant-ph/0102117}{{\tt
  quant-ph/0102117}}].

\bibitem{2013PhRvA..87e4301R}
S.~{Rana}, \emph{{Negative eigenvalues of partial transposition of arbitrary
  bipartite states}},
  \href{http://dx.doi.org/10.1103/PhysRevA.87.054301}{\emph{Phys. Rev. A} {\bf
  87} (May, 2013) 054301}, [\href{https://arxiv.org/abs/1304.6775}{{\tt
  1304.6775}}].

\bibitem{Srednicki}
M.~Srednicki, \emph{Entropy and area},
  \href{http://dx.doi.org/10.1103/PhysRevLett.71.666}{\emph{Phys. Rev. Lett.}
  {\bf 71} (Aug, 1993) 666--669},
  [\href{https://arxiv.org/abs/arXiv:hep-th/9303048}{{\tt
  arXiv:hep-th/9303048}}].

\bibitem{PhysRevLett.84.2726}
R.~Simon, \emph{Peres-horodecki separability criterion for continuous variable
  systems}, \href{http://dx.doi.org/10.1103/PhysRevLett.84.2726}{\emph{Phys.
  Rev. Lett.} {\bf 84} (Mar, 2000) 2726--2729},
  [\href{https://arxiv.org/abs/arXiv:quant-ph/9909044}{{\tt
  arXiv:quant-ph/9909044}}].

\bibitem{PhysRevLett.86.3658}
R.~F. Werner and M.~M. Wolf, \emph{Bound entangled gaussian states},
  \href{http://dx.doi.org/10.1103/PhysRevLett.86.3658}{\emph{Phys. Rev. Lett.}
  {\bf 86} (Apr, 2001) 3658--3661},
  [\href{https://arxiv.org/abs/arXiv:quant-ph/0009118}{{\tt
  arXiv:quant-ph/0009118}}].

\bibitem{PhysRevLett.95.230502}
E.~Shchukin and W.~Vogel, \emph{Inseparability criteria for continuous
  bipartite quantum states},
  \href{http://dx.doi.org/10.1103/PhysRevLett.95.230502}{\emph{Phys. Rev.
  Lett.} {\bf 95} (Nov, 2005) 230502},
  [\href{https://arxiv.org/abs/arXiv:quant-ph/0508132}{{\tt
  arXiv:quant-ph/0508132}}].

\bibitem{BTZ}
M.~Ba\~nados, C.~Teitelboim and J.~Zanelli, \emph{Black hole in
  three-dimensional spacetime},
  \href{http://dx.doi.org/10.1103/PhysRevLett.69.1849}{\emph{Phys. Rev. Lett.}
  {\bf 69} (Sep, 1992) 1849--1851},
  [\href{https://arxiv.org/abs/arXiv:hep-th/9204099}{{\tt
  arXiv:hep-th/9204099}}].

\bibitem{1512.04993}
A.~R. Brown, D.~A. Roberts, L.~Susskind, B.~Swingle and Y.~Zhao,
  \emph{Complexity, action, and black holes},
  \href{http://dx.doi.org/10.1103/PhysRevD.93.086006}{\emph{Phys. Rev. D} {\bf
  93} (Apr, 2016) 086006}, [\href{https://arxiv.org/abs/arXiv:1512.04993}{{\tt
  arXiv:1512.04993}}].

\bibitem{XLQ2016}
P.~Hosur, X.-L. Qi, D.~A. Roberts and B.~Yoshida, \emph{Chaos in quantum
  channels}, \href{http://dx.doi.org/10.1007/JHEP02(2016)004}{\emph{Journal of
  High Energy Physics} {\bf 2016} (Feb, 2016) 4},
  [\href{https://arxiv.org/abs/arXiv:1511.04021}{{\tt arXiv:1511.04021}}].

\bibitem{lennystretch}
L.~Susskind, L.~Thorlacius and J.~Uglum, \emph{The stretched horizon and black
  hole complementarity},
  \href{http://dx.doi.org/10.1103/PhysRevD.48.3743}{\emph{Phys. Rev. D} {\bf
  48} (Oct, 1993) 3743--3761},
  [\href{https://arxiv.org/abs/arXiv:hep-th/9306069}{{\tt
  arXiv:hep-th/9306069}}].

\bibitem{headrick}
M.~Freedman and M.~Headrick, \emph{Bit threads and holographic entanglement},
  \href{http://dx.doi.org/10.1007/s00220-016-2796-3}{\emph{Communications in
  Mathematical Physics} {\bf 352} (May, 2017) 407--438},
  [\href{https://arxiv.org/abs/arXiv:1604.00354}{{\tt arXiv:1604.00354}}].

\bibitem{hubeny}
M.~Headrick and V.~E. Hubeny, \emph{Riemannian and lorentzian flow-cut
  theorems},  \href{https://arxiv.org/abs/arXiv:1710.09516}{{\tt
  arXiv:1710.09516}}.

\bibitem{mmi}
P.~Hayden, M.~Headrick and A.~Maloney, \emph{Holographic mutual information is
  monogamous}, \href{http://dx.doi.org/10.1103/PhysRevD.87.046003}{\emph{Phys.
  Rev. D} {\bf 87} (Feb, 2013) 046003},
  [\href{https://arxiv.org/abs/arXiv:1107.2940}{{\tt arXiv:1107.2940}}].

\bibitem{chaos}
P.~Hosur, X.-L. Qi, D.~A. Roberts and B.~Yoshida, \emph{Chaos in quantum
  channels}, \href{http://dx.doi.org/10.1007/JHEP02(2016)004}{\emph{Journal of
  High Energy Physics} {\bf 2016} (Feb, 2016) 4},
  [\href{https://arxiv.org/abs/arXiv:1511.04021}{{\tt arXiv:1511.04021}}].

\bibitem{cooperative}
V.~E. Hubeny, \emph{Bulk locality and cooperative flows},
  \href{https://arxiv.org/abs/arXiv:1808.05313}{{\tt arXiv:1808.05313}}.

\end{thebibliography}
\end{document}